\documentclass[numberedappendix,twocolumn]{openjournal}

\usepackage{xcolor}
\usepackage{textgreek}
\usepackage[utf8]{inputenc}
\usepackage[english]{babel}

\usepackage{hyperref}
\hypersetup{
    unicode, 
    colorlinks=true,
    linkcolor=linkcolor,
    citecolor=linkcolor,
    filecolor=linkcolor,
    urlcolor=linkcolor,
}
\usepackage{color,colortbl}
\definecolor{linkcolor}{rgb}{0.0,0.3,0.5}
\usepackage{tensind}
\tensordelimiter{?}
\DeclareGraphicsExtensions{.bmp,.png,.jpg,.pdf}
\usepackage{verbatim}
\usepackage[normalem]{ulem}
\usepackage{orcidlink}
\usepackage{soul}
\usepackage{amsmath}

\graphicspath{ {./figs/} }

\addto\extrasenglish{%

}

\defcitealias{Krumholz18a}{Paper I}

\usepackage{threeparttable}

\newcommand{\added}[1]{{#1}}

\begin{document}
\title{Metallicity fluctuation statistics in the interstellar medium and young stars -- II. Elemental cross-correlations and the structure of chemical abundance space}

\author{Mark R. Krumholz\orcidlink{0000-0003-3893-854X}}
\email{mark.krumholz@anu.edu.au}
\affiliation{Research School of Astronomy and Astrophysics, Australian National University, 233 Mt Stromlo Road, Stromlo ACT 2611, Australia}

\author{Yuan-Sen Ting\orcidlink{0000-0001-5082-9536}}
\email{ting.74@osu.edu}
\affiliation{Department of Astronomy, The Ohio State University, Columbus, OH 43210, USA}
\affiliation{Center for Cosmology and AstroParticle Physics (CCAPP), The Ohio State University, Columbus, OH 43210, USA}

\author{Zefeng Li\orcidlink{0000-0001-7373-3115}}
\email{zefeng.li@durham.ac.uk}
\affiliation{Centre for Extragalactic Astronomy, Department of Physics, Durham University, South Road, Durham DH1 3LE, UK}

\author{Chuhan Zhang\orcidlink{0009-0000-2503-4803}}
\email{chuhan.zhang@anu.edu.au}
\affiliation{Research School of Astronomy and Astrophysics, Australian National University, 233 Mt Stromlo Road, Stromlo ACT 2611, Australia}

\author{Jennifer Mead\orcidlink{0009-0006-4744-2350}}
\email{jennifer.mead@columbia.edu}
\affiliation{Department of Astronomy, Columbia University, New York, NY 10027, USA}

\author{Melissa K. Ness\orcidlink{0000-0001-5082-6693}}
\email{melissa.ness@anu.edu.au}
\affiliation{Research School of Astronomy and Astrophysics, Australian National University, 233 Mt Stromlo Road, Stromlo ACT 2611, Australia}
\affiliation{Department of Astronomy, Columbia University, New York, NY 10027, USA}

\begin{abstract}
    Observations of Milky Way stars by multiplexed spectroscopic instruments and of gas in nearby galaxies using integral field units have made it possible to measure the abundances of multiple elements in both the interstellar medium and the stars that form out of it. These observations have revealed complex correlations between elemental abundances, but thus far there has been no \added{analytic} theoretical framework to interpret these data. In this paper we extend the simple stochastically-forced diffusion model of \citet{Krumholz18a}, which has proven successful at explaining the spatial abundance patterns of single elements, to multiple elements, clarifying why elements are correlated and what controls their degree of correlation, and making quantitative predictions for the degree of correlation in both gas and young stars. We show that our results are qualitatively consistent with observed patterns, and point out how application of this theory to measured correlations should enable determination of currently unknown parameters describing r-process nucleosynthesis.
\end{abstract}


\maketitle

\section{Introduction}

The abundance distributions of chemical elements within galaxies represent a vital tracer of the history of star formation, feedback, and nucleosynthesis. These distributions are influenced by, and thus in principle encode, a wealth of details about both the host galaxies and the individual elements: for how long has the galaxy been forming stars? What fraction of the elements produced by those stars are retained in the galaxy? Has the galaxy undergone interactions that mixed or distorted its abundance patterns? Which elements originate from which nucleosynthetic proceses, over what timescales?

The data required to address these questions have become increasingly available thanks to the advent of integral field unit (IFU) measurements of the interstellar medium (ISM) and massively multiplexed spectroscopic measurements of stellar abundances. The former have made it possible to measure not just mean metallicities or radial gradients of the ISM, but full two-dimensional maps.\footnote{A brief note here on nomenclature: in the ISM and extragalactic astronomy communities, the term ``metallicity'' is used to mean the abundance of any heavy element relative to hydrogen, most commonly [O/H], while in the stellar community the term metallicity is used exclusively for the abundance of iron, [Fe/H], or occasionally other chemical metals. Given the background of this paper in the ISM and extragalactic communities, we will generally follow the former convention and use the term metallicity to refer generically to the abundances of all heavy elements. However, when we refer specifically to measurements in stars, we will conform to the stellar community convention and use the alternative term ``abundances'' to make clear that we are not referring exclusively to iron.} The first such maps for oxygen (the element most easily measured in the ISM) began to appear in the last decade \citep[e.g.,][]{Rosales-Ortega11a, Sanders12a, Bresolin15a, Sanchez-Menguiano16a}, and in the past five years a number of authors have analyzed the statistical properties of two-dimensional oxygen abundances maps \citep{Kreckel19a, Kreckel20a, Li21a, Li23a, Metha21a, Metha22a, Metha24a, Williams22a, Gillman22a, Chen24a, Li25a, Myszka25a}. These studies have also begun to extend to other elements, most notably nitrogen. Several authors have measured the statistics of the N/O ratio in pencil-beam surveys \citep{Berg15a, Arellano-Cordova16a, Arellano-Cordova24a, Kumari18a, Dominguez-Guzman22a}, and in the last year these studies have been extended to full nitrogen and sulfur abundance maps analogous to the oxygen ones \citep{Bresolin25a, Li25c}.

Similarly on the stellar side, the literature now contains spectroscopic surveys of up to $\sim 10^6$ Milky Way stars capable of measuring abundances for tens of elements \citep{Majewski17a, Matteucci21a, Buder21a, Buder25a, Meszaros25a}. With data sets of this size and dimensionality, it has become possible to examine the structure of chemical abundance space as recorded on stellar surfaces \citep[e.g.,][]{Weinberg19a, Weinberg22a, Ting22a, Ness22a, Griffith24b, Mead25a}. These studies have revealed that chemical element space is highly structured, with a relatively low effective dimensionality.

Theoretical and simulation modeling efforts to make sense of these observations have begun, but thus far have been limited by resolution. A number of cosmological simulations that have examined the statistical properties of two-dimensional element distributions within galaxies (i.e., going beyond just measuring gradients) or in young stars \citep[e.g.,][]{Grand17a, Orr23a, Bhattarai24a, Li24a}. While these have the advantage of \added{being able to simulate a full Hubble time of evolution}, at their typical mass resolutions of $\sim 0.5$–$1\times 10^4$ M$_\odot$ these simulations \added{reach only $\sim 100$ pc spatial resolution for gas at Milky Way-like mean ISM densities of $\sim 1$ cm$^{-3}$ \citep{Boulares90a}}; this has proven sufficient to capture the $\sim$kpc-scale correlations observed in galaxy metal fields \citep[e.g.,][]{Li24a}, but is clearly insufficient to say much about \added{why, for example, $\sim 1$ pc-sized star clusters appear to have extraordinarily uniform abundances, but there is much less homogeneity on larger scales} \citep[e.g.,][]{Feng14a, Armillotta18a}. Higher-resolution isolated galaxy simulations have only just begun to appear in the literature \citep{Kolborg22a, Zhang25a}. More importantly, however, there have been relatively few attempts thus far to provide a fundamental theoretical basis on which the simulation results can be understood, and which can be used to make general predictions extending into regimes beyond that addressed by the existing, limited simulation suite.

In the previous paper in this series (\citealt{Krumholz18a}; hereafter \citetalias{Krumholz18a}), we provided a minimal theoretical model for the spatial statistics of the distribution of a single element. In this model we treat the metal field in a galactic disk as the result of a competition between a stochastic injection process, which adds metals in sharp peaks at random locations and times, and a diffusion process that smooths out the sharp structures produced by injection. While this model is extraordinarily simple, it has the virtue of being analytically solvable, and the primary prediction made by that solution -- the functional form for the two-point correlation function of an elemental abundance field -- has proven to be a very good match to the statistics of both observed metal fields \citep[e.g.,][]{Kreckel20a, Li21a, Li23a, Li25a, Williams22a, Bresolin25a} and those produced in simulations \citep{Li24a, Zhang25a}.

While this model has proven successful, it has major limitations. It applies only to a single element, injected by a single, dominant nucleosynthetic source. It provides no predictions for the correlations between elements, or even for the spatial statistics of single elements (e.g., nitrogen) for which multiple nucleosynthetic channels make non-negligible contributions to the abundance. Our goal in this paper is to remove some of these limitations, by generalizing the model presented in \citetalias{Krumholz18a} to multiple elements with distinct nucleosynthetic channels. We will then explore the implications of this model for the structure of chemical abundance space.

The remainder of this paper is as follows. In \autoref{sec:model}, we first lay out the formalism of our model and then obtain expressions for the correlations of metal fields within it. In \autoref{sec:discussion} we discuss the implications of the model for abundance distributions in the ISM and young stars. We summarize and conclude in \autoref{sec:conclusion}.

\section{Model}
\label{sec:model}

Here we first introduce some general considerations regarding elemental cross-correlations in \autoref{ssec:delay_times} before presenting our basic model for the origin of elemental abundance correlations in \autoref{ssec:model_basics}. We then obtain a formal solution from the model in \autoref{ssec:formal_sol}, which we 
then use to compute correlation functions (\autoref{ssec:correlation_func}), which are our ultimate aim in this section.

\subsection{Delay times and elemental cross-correlations}
\label{ssec:delay_times}

Before attempting to formulate a model, we can begin by asking why we would expect the abundance distributions for different elements to be correlated at all -- that is, why should regions where oxygen (predominantly from core collapse supernovae) is over-abundant relative to the mean at a given galactocentric radius also tend to have an overabundance of nitrogen (predominantly from AGB stars) or iron (predominantly from thermonuclear supernovae)? Indeed, if the locations and times at which these elements are injected into the ISM of a galaxy were completely uncorrelated with one another, we would expect the metal fields these injections generate to be uncorrelated as well. Nonetheless, both observations \citep{Bresolin25a, Li25c, Mead25a} and simulations \citep{Zhang25a} indicate that such correlations are in fact present, which in turn implies that the times and locations at which these disparate elements are added to the ISM must be correlated with one another.

Upon reflection, however, it is not surprising that correlations exist, at least between some groups of elements, because injection of all elements follows star formation, and the delay time after star formation is not all that long even for elements that are not returned by core collapse supernovae. To illustrate this, consider a simple stellar population formed at time $t=0$. We let $f_{X,0}$ be the mass fraction of element $X$ in this stellar population at birth, and for the purposes of this calculation we adopt Solar abundances following \citet{Asplund09a}. As these stars evolve, they begin to return mass to the ISM, first via massive stellar winds and core collapse supernovae, and later through the evolution of super-AGB and AGB stars and thermonuclear supernovae. We let $\psi_X(t)$ be the cumulative mass of element $X$ returned to the ISM per unit mass of stars formed up to a time $t$ after star formation, so that the additional amount of element $X$ produced per unit mass of stars formed after time $t$ is $p_X(t) = \psi_X(t) - f_{X,0}$.\footnote{To understand the difference between $\psi_X$ and $p_X$, note that the mass return $\psi_X$ includes mass that was incorporated into the star at birth. For example, a star that produces no additional iron via nucleosynthesis, but which formed out of gas containing some iron, will still produce a non-zero return rate for iron, $\psi_\mathrm{Fe} > 0$, because the material that it is returning to the ISM contains the same non-zero iron abundance as the gas from which it was initially formed. Subtracting $f_{X,0}$ removes this effect, isolating just the additional material that is synthesized by stars. A corollary of this, however, is that $p_X(t)$ is neither strictly positive nor strictly monotonic, since some elements are consumed rather than synthesized during some phases of stellar evolution, leaving the mass that is returned during these phases less enriched in that element than the mass from which the stars formed.} 

We compute the contribution to $\psi_X(t)$ for core collapse supernovae and (super-)AGB stars using the \texttt{slug} stellar population synthesis code \citep{da-Silva12a, Krumholz15b}. For the purposes of this calculation we disable \texttt{slug}'s stochastic capabilities and consider a fully-sampled IMF; we adopt a \citet{Chabrier05a} stellar initial mass function, mass-dependent stellar lifetimes from the MIST stellar evolution models \citep{Choi16a}, and mass returns for massive star winds plus core collapse supernovae, super-AGB stars, and AGB stars from the Solar-metallicity models of \citet{Sukhbold16a}, \citet{Doherty14a}, and \citet{Karakas16a}, respectively. Since \texttt{slug} does not include thermonuclear supernovae, we supplement these yields by adding a thermonuclear supernova that we compute by adopting the delay time distribution of \citet{Maoz17a} and the yields per explosion from \citet[their N100 model]{Seitenzahl13a}.

\begin{figure}
    \centerline{
    \includegraphics[width=\columnwidth]{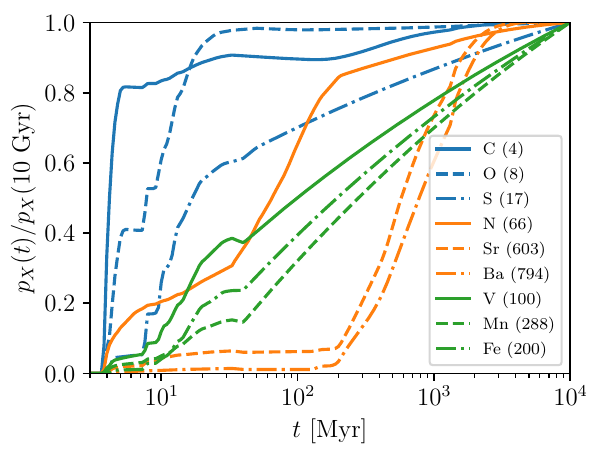}
    }
    \caption{Cumulative distribution function for injection of select elements as a function of time. Lines show the fraction of the element injected at times $<t$ after the formation of a Solar-metallicity simple stellar population, normalized to the production at 10 Gyr, for elements indicated as indicated in the legend; lines are colored by predominant nucleosynthetic source: massive stars (blue), AGB stars (orange), and thermonuclear supernovae (green). Numbers given in the legend indicate the time at which production reaches 50\% of total in units of Myr.\\
    \label{fig:delay_time}
    }
\end{figure}

We plot the resulting values of $p_X(t)$ normalized to the values at $t=10$ Gyr in \autoref{fig:delay_time} for several sample elements. The key point to take from the figure is that even for the elements with the longest delays between star formation and production (e.g., Ba and Sr), most of the production occurs within $\approx 0.5-0.8$ Gyr of star formation, and for most elements, including most of those produced by thermonuclear supernovae and some produced by AGB stars (e.g., N), the delay is considerably smaller. An implication of these timescales is that the stars responsible for nucleosynthesis will not have drifted that far (in the frame co-rotating with the galaxy) between formation and element return. For example, if we consider two stars whose velocities differ by $\sim 2$ km s$^{-1}$ at birth (typical of the velocity differences inferred in young dispersing clusters -- e.g., \citealt{Crundall19a, Zerjal23a, Armstrong24a}), then over the $\sim 50$ Myr gap between when the bulk of O and the bulk of N are injected, these stars will drift apart by only $\sim 100$ pc, and even the $\sim 250$ Myr gap between O and Mn only corresponds to $\sim 500$ pc of drift. These figures are significantly less than the $\sim 2$ kpc correlation lengths of the metal fields seen in nearby Milky Way-like galaxies \citep[e.g.,][]{Li21a, Li23a}.

The basic conclusion of this analysis is that we should approximate the injection of most elements as occurring at sites that are correlated in space and time. The nature of this correlation will in turn be crucial to determining the correlation between the resulting elemental abundances in the ISM, and thence in young stars. This forms the basic premise for the model we now proceed to describe.

\subsection{Description of the model}
\label{ssec:model_basics}

\begin{table*}
    \begin{center}
    \begin{threeparttable}
    \caption{\added{Parameters of the model}
    \label{tab:parameters}}
    \begin{tabular}{llll}
        \hline\hline
        Parameter & Meaning & Definition & Fiducial value\tnote{1} \\ \hline
        $\kappa$ & ISM diffusion coefficient & \autoref{eq:evoleq} & 350 pc km s$^{-1}$
        \\
        $\Gamma$ & Rate of star formation events & \autoref{ssec:model_basics}, \autoref{item:gamma_def} & 3 Myr$^{-1}$ kpc$^{-2}$ \\
        $\sigma_{M_*}^2$ & Variance in mass of star-forming events & \autoref{ssec:model_basics}, \autoref{item:sigmaMstar} & -- \\
        $\sigma_v$ & Velocity dispersion for stellar drift & \autoref{ssec:model_basics}, \autoref{item:sigmav} & 2 km s$^{-1}$ \\
        $\Delta t_{X,i}$ & Delay for return of element $X$ by channel i & \autoref{ssec:model_basics}, \autoref{item:yields} & - \\
        $y_{X,i}$ & Yield of element $X$ via channel $i$ & \autoref{ssec:model_basics}, \autoref{item:yields} & Element-dependent (\autoref{tab:delay_times}) \\
        $\sigma_i$ & Spatial dispersion of element return by channel $i$ & \autoref{ssec:model_basics}, \autoref{item:dispersion} &
        60 pc (SNe), 1 pc (AGBs) \\
        $f_{X,i}$ & Fraction of element $X$ produced by channel $i$ & \autoref{eq:fdefn} & Element-dependent (\autoref{tab:delay_times}) \\
        $\sigma_w^2$ & Non-dimensionalized version of $\sigma_{M_*}^2$ & \autoref{ssec:formal_sol} & 120 \\
        $\ell_\mathrm{corr}$ & Spatial correlation length of metal field & \autoref{ssec:correlation_func} & 1 kpc \\
        $\ell_{X,i}$ & Diffusion length corresponding to $\Delta t_{X,i}$ & \autoref{ssec:correlation_func} & Set by $\kappa$, $\Delta t_{X,i}$ (\autoref{eq:lXj}) \\
        $\ell_{XY,ij}$ & Drift length corresponding to $\Delta t_{X,i}-\Delta t_{Y,j}$ & \autoref{ssec:correlation_func} & Set by $\sigma_v$, $t_{X,i}$, $t_{Y,j}$ (\autoref{eq:lXYij})
        \\ \hline
    \end{tabular}
    \begin{tablenotes}
        \item[1] \added{Fiducial values are provided only for quantities that appear in the final numerical result.}
    \end{tablenotes}
    \end{threeparttable}
    \end{center}
\end{table*}

Our model here is a straightforward extension of the one presented in \citetalias{Krumholz18a}\added{, and we summarize the parameters appearing in this model in \autoref{tab:parameters}}. We describe the distribution of some element $X$ in a galaxy in terms of two-dimensional surface density $\Sigma_X$, which is a function of position $\mathbf{x}$ and time $t$. We assume that the metal field is set by a combination of diffusion and injection, so that the evolution of $\Sigma_X$ obeys
\begin{equation}
    \frac{\partial}{\partial t}\Sigma_X = \kappa \nabla^2\Sigma_X + \dot{\Sigma}_{X,\mathrm{src}},
    \label{eq:evoleq}
\end{equation}
where $\kappa$ is a constant diffusion coefficient and $\dot{\Sigma}_{X,\mathrm{src}}$ is a stochastic source term describing the injection of metals by stellar mass loss. The system has open spatial boundaries (i.e., $\Sigma_X$ is defined at all positions $\mathbf{x}$ in an infinite 2d plane), and the initial condition is $\Sigma_X = 0$ at all positions at $t=0$.

Most of the scientific content of the model is in the choice of source term $\dot{\Sigma}_{X,\mathrm{src}}$. In \citetalias{Krumholz18a}, we treat the source term as Poissonian in space and time, neglecting time delays between star formation and element injection; here we make a minimal modification to this approach to explicitly account for delays that may differ from one element to another. To this end, we assume
\begin{enumerate}
    \item Stars form only at positions $\mathbf{x}$ such that $|\mathbf{x}| < L$. We will always evaluate all quantities in the limit $L\to\infty$, so that we are approximating the galaxy as an infinite two-dimensional plane. \added{Conceptually $L$ is just a dummy variable, and we can think of taking the limit as $L\to\infty$ as implying that we will be considering the behavior of the metal field only on spatial scales much smaller than the characteristic size of the galaxy as whole.}
    \item Within the star-forming region, star formation events (where ``event'' refers to formation of a stellar cluster or association, not a single star) occur at some rate per unit time per unit area $\Gamma$. The expected number of star formation events up to some time $t$ is therefore $\langle N_\mathrm{sf}\rangle = \pi L^2 \Gamma t$. \label{item:gamma_def}
    \item Star formation is a Poissonian process, so that for any given realization of the stochastic source term, the number of events $N_\mathrm{sf}$ is drawn from a Poisson distribution with expectation value $\langle N_\mathrm{sf}\rangle$, and the locations and times of those injection events are drawn from independent uniform distributions within the allowed range of positions $|\mathbf{x}| < L$ and time $(0,t)$. 
    \item Each star formation event produces a mass $M_*$ of stars, which may also be a random variable drawn from a distribution; we require that this distribution be independent of the distributions for number of star formation events, position, and time, and that it have finite expectation value $\langle M_*\rangle$ and variance $\sigma_{M_*}^2$, but place no other conditions on it. Note that we do \textit{not} require that $M_*$ be drawn from a Gaussian distribution, simply that the distribution from which it is drawn has finite variance.
    \label{item:sigmaMstar}
    \item The stars that form drift at a constant two-dimensional random velocity $\mathbf{v}$ relative to the background gas and metal field. For simplicity we will assume that $\mathbf{v}$ is chosen from a two-dimensional Gaussian distribution with fixed dispersion $\sigma_v$.
    \label{item:sigmav}
    \item At fixed times $\Delta t_{X,1}, \Delta t_{X,2}, \ldots$ after star formation, the formed stars return masses $m_{X,1} = y_{X,1} M_*, m_{X,2} = y_{X,2} M_*, \ldots$ of element $X$ to the gaseous metal field, where $y_{X,j}$ is a fixed yield fraction; \added{thus we are approximating the process of mass return by stars as occurring in a series of discrete bursts at fixed times after star formation, with each burst consisting of a mass of each metal that is directly proportional to the mass of stars formed, with no stochastic variation. We will envision these bursts as corresponding to different nucleosynthetic channels, each of which is approximated as occurring at a single time. Thus for example the first burst will represent return of all metals produced by core collapse supernovae, the second return of all metals produced by thermonuclear supernovae, the third return of all metals by AGB stars, and so forth, and we will generically refer to these different bursts as nucleosynthetic channels. We discuss future ways of improving this rather crude approximation in \autoref{ssec:caveats}.} \label{item:yields}
    \item When metals are injected, they are deposited with a distribution that is an isotropic Gaussian in space, centered on the present position of the stars, with some fixed dispersion $\sigma_1, \sigma_2,\ldots$; conceptually, this represents the characteristic size of the injection produced by a particular nucleosynthetic channel, for example the size of a supernova remnant or the bubble blown by an AGB wind.
    \label{item:dispersion}
\end{enumerate}
Note that these assumptions are largely identical to those made in \citetalias{Krumholz18a}, and reduce to them exactly if we set the random velocity $\sigma_v$ to zero and assume that there is a only a single nucleosynthetic channel with time delay $\Delta t_{X,1} = 0$.

Given these assumptions, we can formally write out the source term as
\begin{eqnarray}
    \lefteqn{
    \dot{\Sigma}_{X,\mathrm{src}} = \sum_{i=1}^{N_\mathrm{sf}} M_{*,i} \sum_{j=1}^{N_\mathrm{ch}} y_{X,j} 
    } \quad
    \nonumber \\
    & &
    \mathcal{N}(\mathbf{x}-\mathbf{x}_{i} - \mathbf{v}_i \Delta t_{X,j}, \sigma_j^2\mathbb{I}) \, \delta(t-t_{i}-\Delta t_{X,j}),
    \label{eq:src_term}
\end{eqnarray}
where $N_\mathrm{sf}$ is the number of star formation events that have occurred, $M_{*,i}$, $\mathbf{x}_i$, $\mathbf{v}_i$ and $t_i$ are the stellar mass formed, position, drift velocity, and time of star formation event $i$, there are $N_\mathrm{ch}$ nucleosynthetic channels\footnote{For notational simplicity we will treat $N_\mathrm{ch}$ as the same for all elements, and account for the fact that some channels do not contribute to some elements simply by setting the yield $y_{X,j} = 0$ for those cases.} with associated delay times $\Delta t_{X,j}$, yields $y_{X,j}$, and injection dispersions $\sigma_{X,j}$, $\mathcal{N}(\boldsymbol{\mu},\mathbb{S})$ is the standard two-dimensional normal distribution with central value $\boldsymbol{\mu}$ and covariance matrix $\mathbb{S}$, and $\mathbb{I}$ is the identity matrix. The quantities $N_\mathrm{sf}$, $\mathbf{x}_i$, $\mathbf{v}_i$, and $t_i$ are random variables, which are drawn, respectively, from the Poisson distribution $\mathcal{P}_\lambda(N_\mathrm{sf})$ with expectation value $\lambda =\langle N_\mathrm{sf}\rangle$, a two-dimensional distribution that is uniform inside a circle of radius $L$ and zero outside it, a two-dimensional Gaussian distribution with dispersion $\sigma_v$, and a one-dimensional distribution that is uniform for times in $(0,t)$ and zero otherwise. The stellar mass $M_{*,i}$ is also a random variable, which is drawn from a distribution $p_{M_*}(M_{*,i})$ with known expectation value $\langle M_*\rangle$ and variance $\sigma_{M_*}^2$.

\added{Before moving on to predict the statistics of elemental cross-correlations in the context of this model, it is important to enumerate some of the simplifications of our model. While these are necessary for analytic tractability, it is important to understand where potential problems may arise, and we will discuss the potential impact of some of these simplifications further in \autoref{ssec:caveats}. For the moment, we list here for reader convenience the major omissions in our model:}
\begin{enumerate}
    \item \added{We treat star formation as a Poisson process in space and time, when in reality it is correlated on many scales -- on large scales by the presence of galactic-scale features such as spiral arms and bars, and on smaller scales by the fractal structure of turbulent gas \citep[e.g.,][]{Grasha17a, Grasha17b, Menon21a}. This structure will change both where metals are injected and the metallicity distributions in the stars that form relative to our simple assumption.}
    \item \added{We neglect the finite spread in element return times, and instead approximate that each return channel for each element injects metals at a single time.}
    \item \added{We assume that the initial injection of elements from each return channel takes the form of an isotropic Gaussian, whose width the the same for all elements returned via a given channel (e.g., all elements returned as a result of core collapse supernovae are initially deposited with a Gaussian profile of fixed width).}
    \item \added{We neglect loss of metals from the interstellar medium via galactic winds. Uniform metal removal (i.e., removal that leaves abundances of elements in the remaining gas unchanged), as is commonly assumed in cosmological simulations, would not affect the correlations in which we are interested. However, there is strong numerical \citep[e.g.,][]{Vijayan24a, Vijayan25a}, observational \citep[e.g.,][]{Martin02a, Chisholm18a, Lopez20a, Lopez23a, Cameron21a}, and circumstantial \citep[e.g.,][]{Peeples14a, Forbes19a, Sharda21b} evidence that galactic winds are metal-loaded and preferentially eject some elements, which could affect abundance ratios.}
    \item \added{We treat both diffusion of metals and drift of stars as isotropic Gaussian processes, ignoring the non-Gaussian nature of turbulent mixing \citep[e.g.,][]{Colbrook17a} and the presence of large-scale coherent mixing processes such as those driven by spiral arms \citep[e.g.,][]{Yang12a, Petit15a}.}
\end{enumerate}

\subsection{Formal solution}
\label{ssec:formal_sol}

With the model now specified, we proceed to non-dimensionalize and obtain a formal solution following \citetalias{Krumholz18a}. Our non-dimensional version of the evolution equation (\autoref{eq:evoleq}) is
\begin{eqnarray}
    \lefteqn{
    \frac{\partial}{\partial \tau} S_X = \nabla_r^2 S_X + \sum_{i=1}^{N_\mathrm{sf}} w_i \sum_{j=1}^{N_\mathrm{ch}} f_{X,j}
    } \quad
    \nonumber \\
    & &
    \mathcal{N}(\mathbf{r} - \mathbf{r}_i - \mathbf{u}_i\Delta\tau_{X,j}, s_{j} \mathbb{I}) \,\delta(\tau-\tau_i - \Delta \tau_{X,j}),
    \label{eq:evoleqnd}
\end{eqnarray}
where $\nabla_r$ indicates differentiation with respect to the dimensionless position $\mathbf{r}$, and we have introduced the dimensionless time, position, velocity, metal surface density, and metal mass, respectively:
\begin{eqnarray}
    \tau & = & t \sqrt{\Gamma\kappa}
    \\
    \mathbf{r} & = & \mathbf{x} \left(\frac{\Gamma}{\kappa}\right)^{1/4} 
    \\
    \mathbf{u} & = & \mathbf{v} \left(\frac{1}{\Gamma\kappa^3}\right)^{1/4}
    \\
    S_X & = & \Sigma_X \frac{\kappa^{1/2}}{\Gamma^{1/2} \langle m_X\rangle}
    \\
    w_i & = & \frac{m_{X,i}}{\langle m_X\rangle}.
\end{eqnarray}
In the expressions above, $s_j$ is related to $\sigma_j$ as $\mathbf{r}$ is to $\mathbf{x}$, and we have introduced the expected metal mass per star formation event returned by all channels and the fractional contribution from each channel, defined respectively by
\begin{eqnarray}
    \langle m_X \rangle & = & \sum_{j=1}^{N_\mathrm{ch}} y_{X,j} \langle M_*\rangle
    \\
    f_{X,j} & = & \frac{y_{X,j}}{\sum_{j=1}^{N_\mathrm{ch}} y_{X,j}}.
    \label{eq:fdefn}
\end{eqnarray}

One can immediately verify that \autoref{eq:evoleqnd} has the exact solution
\begin{equation}
    S_X = \sum_{i=1}^{N_\mathrm{sf}} w_i \sum_{j=1}^{N_\mathrm{ch}} \mathcal{N}\left(\mathbf{r}-\mathbf{p}_{X,ij}, [2\theta_{X,ij} + s_{j}^2]\mathbb{I}\right)
    H(\theta_{X,ij}),
    \label{eq:formalsol}
\end{equation}
as a function of dimensionless position $\mathbf{r}$ and time $\tau$, where for brevity we have introduced the symbols
\begin{eqnarray}
    \mathbf{p}_{X,ij} = \mathbf{r}_i + \mathbf{u}_i\Delta\tau_{X,j}
    & \quad &
    \theta_{X,ij} = \tau - \tau_i -\Delta\tau_{X,j},
\end{eqnarray}
and $H(x)$ is the Heaviside step function, which is zero for negative arguments and unity for positive arguments. This solution is identical to that derived for the formal system in \citetalias{Krumholz18a}, with the exception of the $-\Delta \tau_{X,j}$ terms inside the step function, which serve to introduce the time delay between star formation and element return, and the $-\mathbf{u}_i \Delta\tau_{X,j}$ terms inside the Gaussian, which serve to introduce the spatial displacement associated with that delay.

In this formal solution, the number of star formation events $N_\mathrm{sf}$ is drawn from a Poisson distribution with expectation value $\langle N_\mathrm{sf}\rangle = \pi R^2 \tau$. The dimensionless random variables $\mathbf{r}_i$, $\mathbf{u}_i$, and $\tau_i$ are drawn from the distributions
\begin{eqnarray}
    p_\mathbf{r}(\mathbf{r}') & = & \frac{1}{\pi R^2} H(R - |\mathbf{r}'|) \\
    p_\mathbf{u}(\mathbf{u}') & = & \frac{1}{2\pi \sigma_u^2} e^{-u'^2/2\sigma_u^2}
    \\
    p_\tau(\tau') & = & \frac{1}{\tau} H(\tau') H(\tau-\tau'),
\end{eqnarray}
where $R = L (\Gamma/\kappa)^{1/4}$, $\sigma_u = \sigma_v (1/\Gamma\kappa^3)^{1/4}$. Note that the Heaviside functions in $p_\mathbf{r}$ and $p_\tau$ serve to confine the positions and times of star formation events to lie within a circle of radius $R$ and at times from 0 to $\tau$, respectively. Finally, the injection masses $w_i$ are drawn from a distribution $p_w(w')$ which has the same shape as $p_{M_*}$, but which has been rescaled to have expectation value $\langle w\rangle = 1$ and variance $\sigma_w^2 = \sigma_{m_X}^2 / \langle m_X\rangle^2$. 

We are interested in the expected statistical properties of the metal fields $S_X$ generated by different random realizations of the stochastic variables. For this purpose, we define the expectation value of any quantity $q$ by
\begin{eqnarray}
    \langle q \rangle & \equiv &
    \lim_{R\to\infty} \frac{1}{\pi R^2\tau} \sum_{N_\mathrm{sf} = 0}^\infty 
    \mathcal{P}_\lambda(N_\mathrm{sf}) \int_0^\infty p_w(w) \int p_\mathbf{u}(\mathbf{u}')
    \nonumber \\
    & & \quad
    \int_0^\tau 
    \int_{\mathbf{r}'<R} q(N_\mathrm{sf}, w, \mathbf{r'},\mathbf{u}' \tau') \, d\mathbf{r}' \, d\tau' \, d\mathbf{u}'\, dw.
    \label{eq:expect_defn}
\end{eqnarray}
In evaluating expressions of this form we shall always assume that $\tau>\Delta\tau_{X,j}$ for all $j$, i.e., star formation has been ongoing long enough that there has been time for every nucleosynthetic channel to contribute for each element.

From the formal solution and the definition of the expectation value, we can immediately compute the expected dimensionless metallicity $\langle S_X\rangle$. Evaluation of the integrals in \autoref{eq:expect_defn} is straightforward because each of the random variables is drawn independently and the integrals are therefore separable. Carrying out the algebra, we obtain
\begin{eqnarray}
    \langle S_X\rangle = \tau - \sum_{j=1}^{N_\mathrm{ch}} f_{X,j} \Delta\tau_{X,j}.
    \label{eq:SX_avg}
\end{eqnarray}
This has a simple interpretation: in our dimensionless units, we expect one star formation event per unit area per unit time, which eventually injects one unit of mass. Thus after a long time we should expect $\langle S_X\rangle$ to approach $\tau$. However, stars that formed less than a time $\Delta\tau_{X,j}$ ago have not yet provided their metal injection through channel $j$, and this correspondingly reduces the expected value of $S_X$.

\subsection{Correlation functions}
\label{ssec:correlation_func}

We now seek to evaluate the expected cross-correlation between the metal fields $S_X$ and $S_Y$ for two elements $X$ and $Y$ that are produced by the same set of star formation events but with different delay times and different fractional contributions from each channel. We define the cross-correlation in the usual way, as
\begin{equation}
    \xi_{XY}(\mathbf{r}) = \left\langle\frac{ \overline{\left[S_X(\mathbf{r} + \mathbf{r}')-\overline{S_X}\right] \left[S_Y(\mathbf{r}')-\overline{S_Y}\right]}}{\sigma_{r,S_X} \sigma_{r,S_Y}}\right\rangle,
    \label{eq:crosscorr_defn}
\end{equation}
where the overline indicates averaging over space, i.e., for any quantity $q$ we define
\begin{equation}
    \overline{q} = \lim_{R\to\infty} \frac{1}{\pi R^2} \int_{\mathbf{r}'<R} q(\mathbf{r}') \, d\mathbf{r}'.
\end{equation}
We similarly define $\sigma^2_r$ as the variance of the metal field over position, i.e.,
\begin{equation}
    \sigma^2_{r,S_X} \equiv \lim_{R\to\infty} \frac{1}{\pi R^2} \int_{\mathbf{r}'<R} \left[ S_X(\mathbf{r}') - \overline{S_X}\right]^2 \, d\mathbf{r}',
\end{equation}
and similarly for $S_Y$. However, because $S_X$ is statistically uniform across all space (in the limit $R\to\infty$), we immediately have $\overline{S_X} = \langle S_X\rangle$ and $\sigma_{r,S_X} = \sigma_{S_X}$, where
\begin{equation}
    \sigma^2_{S_X} = \left\langle S_X^2\right\rangle - \left\langle S_X\right\rangle^2
    \label{eq:variance_defn}
\end{equation}
is the variance in $S_X$ at any given position over all realizations of the random variables; analogous expressions hold for $S_Y$. Thus the cross-correlation immediately reduces to
\begin{equation}
    \xi_{XY}(\mathbf{r}) = \frac{\left\langle \overline{S_X(\mathbf{r} + \mathbf{r}') S_Y(\mathbf{r}')}\right\rangle - \langle S_X\rangle\langle S_Y\rangle}{\sigma_{S_X} \sigma_{S_Y}}.
    \label{eq:crosscorr_reduced}
\end{equation}

Our remaining task is to evaluate the first term in the numerator; once we have done so, we can then immediately deduce the variances $\sigma_{S_X}$ and $\sigma_{S_Y}$ in the denominator from the requirement that at zero lag every element cross-correlates perfectly with itself, i.e., $\xi_{XX}(0) = 1$. In order to evaluate this term, we invoke the cross-correlation theorem, which implies
\begin{eqnarray}
    \left\langle \overline{S_X(\mathbf{r}+\mathbf{r}') S_Y(\mathbf{r}')}\right\rangle
    & = & \lim_{R\to\infty} \frac{1}{\pi R^2} \int \left\langle S_X(\mathbf{r}+\mathbf{r}') S_Y(\mathbf{r}') \right\rangle \, d\mathbf{r}' 
    \nonumber \\
    & = &
    2\pi \int_0^\infty \Psi_{XY}(k) J_0(kr) k \, dk,
    \label{eq:crosscorr_thm}
\end{eqnarray}
where $J_n$ is the Bessel function of the first kind of order $n$,
\begin{equation}
    \Psi_{XY}(k) = \lim_{R\to\infty} \frac{1}{\pi R^2} \left\langle \tilde{S}^*_X(\mathbf{k}) \tilde{S}_Y(\mathbf{k}) \right\rangle
    \label{eq:crosscorr_spec}
\end{equation}
is the cross-correlation power spectrum, and $\tilde{S}_X(\mathbf{k})$ is the Fourier transform of metal field,
\begin{eqnarray}
    \tilde{S}_{X}(\mathbf{k}) & = & \frac{e^{-\tau k^2}}{2\pi}
    \sum_{i=1}^{N_\mathrm{sf}} w_i e^{\tau_i k^2 + i\mathbf{k}\cdot\mathbf{r}_i}
    \sum_{j=1}^{N_\mathrm{ch}} f_{X,j}
    \nonumber \\
    & &
    \quad
    e^{i \mathbf{k}\cdot \mathbf{u}_i \Delta\tau_{X,j}}
    e^{(\Delta\tau_{X,j}-s_{j}^2/2)k^2}
    H(\theta_{X,ij}).
    \label{eq:fourier}
\end{eqnarray}
Note that in expressing $\Psi_{XY}$ as a function solely of the magnitude $k=|\mathbf{k}|$ of the vector wavenumber $\mathbf{k}$, and equivalently writing the correlation as a function of the scalar separation $r=|\mathbf{r}|$, we have relied on the symmetry of our model to conclude that the expected metal field is statistically isotropic.

From this point evaluating the power spectrum $\Psi_{XY}$ is simply a matter of evaluating expectation values of the form given by \autoref{eq:expect_defn}. This is straightforward but algebraically somewhat lengthy, so we defer the calculation to \autoref{app:crosscorr_eval}. The final outputs of this calculation are the cross-correlation function $\xi_{XY}(r)$ (\autoref{eq:crosscorr_final}) and the cross-correlation at zero lag $\Xi_{XY} \equiv \xi_{XY}(r=0)$ (\autoref{eq:crosscorr_zero_final}). The latter is of particular interest because this is what will define the correlation of elements in stars, which will record the abundances of the interstellar gas at the point in space and time where they form.

To re-express the dimensionless results derived in \autoref{app:crosscorr_eval} in terms of dimensional variables, we follow \citet{Li21a} in defining the correlation length as $\ell_\mathrm{corr} \equiv \sqrt{\kappa t}$; by analogy we also define the length scales associated with diffusion over the delay time of a single element,
\begin{equation}
    \ell_{X,j} = \sqrt{\kappa\Delta t_{X,j}}
    \label{eq:lXj}
\end{equation}
and the length scale over which stars drift between two injection events, 
\begin{equation}
    \ell_{XY,ij} = \sigma_v |\Delta t_{X,i} - \Delta t_{Y,j}|/\sqrt{2}.
    \label{eq:lXYij}
\end{equation} With these definitions, our final cross-correlation function is
\begin{eqnarray}
    \xi_{XY}(x) & = &
    \frac{1+\sigma_w^2}{4\pi \sigma_{S_X}\sigma_{S_Y}}
    \sum_{i=1}^{N_\mathrm{ch}} 
    \sum_{j=1}^{N_\mathrm{ch}}
    f_{X,i} f_{Y,j}
    \nonumber \\
    & &
    \int_0^\infty \frac{J_0(ax)}{a} e^{-(\sigma_{X,i}^2+\sigma_{Y,j}^2)a^2/2}
    \nonumber \\
    & & 
    \quad
    \left(e^{-|\ell_{X,i}^2-\ell_{Y,j}^2|a^2}
    - e^{(\ell_{i}^2+\ell_{j}^2-2\ell_\mathrm{corr}^2)a^2}
    \right)
    \nonumber \\
    & &
    \quad
    \left[
    1 - 2 a \ell_{XY,ij} F\left(a \ell_{XY,ij}\right)
    \right]
    \,
    da,
    \label{eq:crosscorr_dim}
\end{eqnarray}
where $F(x)$ is the Dawson integral, defined by $F(x) = \int_0^x \exp(y^2-x^2)\, dy$, and the cross-correlation at zero lag is
\begin{eqnarray}
    \Xi_{XY} & = & \frac{1+\sigma_w^2}{8\pi\sigma_{S_X}\sigma_{S_Y}}
    \sum_{i=1}^{N_{\mathrm{ch}}}
    \sum_{j=1}^{N_{\mathrm{ch}}}
    f_{X,i} f_{Y,j}
    \nonumber \\
    & &
    \quad
    \left(
    \ln \alpha_{XY,ij}
    +
    2 \sqrt{\frac{\beta_{XY,ij}^2}{1+\beta_{XY,ij}^2}}\sinh^{-1}\beta_{XY,ij}
    \right.
    \nonumber \\
    & &
    \qquad
    \left.
    {} -
    2 \sqrt{\frac{\gamma_{XY,ij}^2}{1+\gamma_{XY,ij}^2}}\sinh^{-1}\gamma_{XY,ij}
    \right).
    \label{eq:crosscorr_zero_dim}
\end{eqnarray}
In these expressions, $\sigma_X^2$ is the variance of metal field $X$, given by
\begin{eqnarray}
    \sigma_{S_X}^2 & = &
    \frac{1+\sigma_w^2}{8\pi}
    \sum_{i=1}^{N_{\mathrm{ch}}}
    \sum_{j=1}^{N_{\mathrm{ch}}}
    f_{X,i} f_{X,j}
    \nonumber \\
    & &
    \quad
    \left(
    \ln \alpha_{XX,ij}
    +
    2 \sqrt{\frac{\beta_{XX,ij}^2}{1+\beta_{XX,ij}^2}}\sinh^{-1}\beta_{XX,ij}
    \right.
    \nonumber \\
    & &
    \qquad
    \left.
    {} -
    2 \sqrt{\frac{\gamma_{XX,ij}^2}{1+\gamma_{XX,ij}^2}}\sinh^{-1}\gamma_{XX,ij}
    \right),
    \label{eq:sigma_sx_dim}
\end{eqnarray}
and the various dimensionless factors are
\begin{eqnarray}
    \alpha_{XY,ij} & = & \frac{4\ell_\mathrm{corr}^2-2\ell_{X,i}^2-2\ell_{Y,j}^2 + \sigma_{i}^2+\sigma_{j}^2}{2 |\ell_{X,i}^2-\ell_{Y,j}^2| + \sigma_i^2 + \sigma_j^2}
    \label{eq:alpha}
    \\
    \beta_{XY,ij} & = & \frac{\ell_{XY,ij}}{\sqrt{2\ell_\mathrm{corr}^2 - \ell_{X,i}^2-\ell_{Y,j}^2 + \sigma_i^2/2 + \sigma_j^2/2}}
    \label{eq:beta}
    \\
    \gamma_{XY,ij} & = & \frac{\ell_{XY,ij}}{\sqrt{|\ell_{X,i}^2-\ell_{Y,j}^2| + \sigma_i^2/2 + \sigma_j^2/2}}.
    \label{eq:gamma}
\end{eqnarray}
This provides our result for the cross-correlation between two elements, and lets us evaluate the expected cross-correlation as a function of the host galaxy correlation length $\ell_\mathrm{corr}$, the diffusion coefficient for the ISM $\kappa$, the time delays and injection widths for each element and production channel $\Delta t_{X,j}$ and $\sigma_j$, and the velocity dispersion for stellar drift $\sigma_v$. \added{Note that our assumption in \autoref{ssec:formal_sol} that we are working at a time $t > \Delta t_{X,i}$ for all elements $X$ and channels $i$ implies that $\ell_\mathrm{corr} > \ell_{X,i}$, and thus that $\alpha_{XY,ij}$ is strictly positive and $\beta_{XY,ij}$ is real.}

Finally, note that in observations the quantity measured is often not the cross-correlation of the metallicity itself, but of the logarithm of metallicity, i.e.,
\begin{eqnarray}
    \lefteqn{
    \xi_{\log X\log Y} =
    }
    \nonumber \\
    & &
    \!\!\!\!\!
    \frac{\left\langle \overline{\log S_X(\mathbf{r}+\mathbf{r}') \log S_Y(\mathbf{r}')}\right\rangle - \left\langle \overline{\log S_X}\right\rangle \left\langle \overline{\log S_Y}\right\rangle}{\sigma_{\log S_X} \sigma_{\log S_Y}},
\end{eqnarray}
where $\sigma^2_{\log S_X}$ is the variance in the logarithm of the metal field over all realizations of the random variables (cf.~\autoref{eq:variance_defn}). While in general the $\xi_{\log X \log Y}$ and $\xi_{XY}$ can be arbitrarily different, in practice they are always similar. The fundamental reason for this is that the Pearson correlation is invariant under linear transformations of the variables, and because the ranges of $S_X$ and $S_Y$ are small in most data sets of interest, the logarithm is very close to a linear transformation. To see this, let the metal field at any point be $S_X = \langle S_X\rangle + \delta S_X$, where $\delta S_X$ is the deviation from the mean metallicity $\langle S_X\rangle$ as a function of position. Then
\begin{eqnarray}
    \log S_X & = &\log \langle S_X\rangle  + \log\left(1 + \frac{\delta S_X}{\langle S_X\rangle}\right)
    \nonumber \\
    & = & \log \langle S_X\rangle + \frac{1}{\ln 10} \frac{\delta S_X}{\langle S_X\rangle} + O\left(\frac{\delta S_X^2}{\langle S_X\rangle^2}\right)
\end{eqnarray}
where in the second step we have Taylor expanded the logarithm about $\delta S_X/\langle S_X\rangle = 0$. This parameter is in practice small: $\langle \delta S_X\rangle^2 = \sigma_{S_X}^2$, and for the realistic parameters we use for numerical evaluations in \autoref{sec:discussion} we find that $\sigma_{S_X}^2/\langle S_X\rangle^2<0.012$ for all elements. We can therefore drop this term, and then make a final substitution to obtain
\begin{equation}
    \log S_X \approx 
    \log \langle S_X\rangle + \frac{1}{\ln 10} \left(\frac{S_X}{\langle S_X\rangle} - 1\right),
\end{equation}
which is manifestly a linear transformation. Thus in practice we need not distinguish between Pearson correlations of the metallicity or its logarithm.

\section{Discussion}
\label{sec:discussion}

We now consider the implications of our findings. In \autoref{ssec:crosscorr_table}, we predict typical correlations between different elements. We then consider how observational effects will cause measured correlations to differ from true ones for both gas (\autoref{ssec:obs_eff_gas}) and stars (\autoref{ssec:obs_eff_stars}). We next compare to existing data sets in \autoref{ssec:obs_comparison} and discuss the limits of our model and potential future improvements to it in \autoref{ssec:caveats}.

\subsection{Expected elemental cross-correlations in gas and mono-age stellar populations}
\label{ssec:crosscorr_table}

Armed with \autoref{eq:crosscorr_zero_dim}, we are now in a position to predict approximate degrees of cross-correlation between different elements for a Milky Way-like galaxy in both gas and mono-age stellar populations. \added{To do so we must adopt numerical values for the various parameters appearing in the model; we summarize our fiducial choices in \autoref{tab:parameters}, and we quantify the sensitivity of our results to those choices in \autoref{app:parameter_sensitivity}.} We adopt a correlation length $\ell_\mathrm{corr}=1$ kpc, typical of values measured for Milky Way-mass galaxies \citep{Li21a, Li23a}, and similarly following \citeauthor{Li21a}~a diffusion coefficient $\kappa = h \sigma_g/3$, where $h\approx 150$ pc and $\sigma_g\approx 7$ km s$^{-1}$ are the approximate scale height and velocity dispersion of neutral gas in the Galaxy \citep{Kalberla09a}. We adopt $\sigma_v = 2$ km s$^{-1}$, a typical velocity dispersion in young, dispersing star clusters \citep[e.g.][]{Crundall19a, Zerjal23a, Armstrong24a}.

\begin{deluxetable}{@{\extracolsep{2pt}}lcccccc}
    \tablecaption{Production fractions and delay times for selected elements
    \label{tab:delay_times}
    }
    \tablehead{
    & \multicolumn{2}{c}{CCSN} & \multicolumn{2}{c}{AGB} & \multicolumn{2}{c}{TNSN} 
    \\ \cline{2-3} \cline{4-5} \cline{6-7}
    \colhead{Z} & \colhead{$f$} & \colhead{$t_\mathrm{50}$} & \colhead{$f$} & \colhead{$t_\mathrm{50}$} & \colhead{$f$} & \colhead{$t_\mathrm{50}$} \\
    & & \colhead{[Myr]} & & \colhead{[Myr]} & & \colhead{[Myr]}
    }
    \startdata
     6 (C) 
& 0.83 & \phantom{00}4.2
& 0.17 & 575.4
& 0.00 & -
\\
 7 (N) 
& 0.32 & \phantom{00}6.0
& 0.68 & 114.8
& 0.00 & -
\\
 8 (O) 
& 0.81 & \phantom{00}7.9
& 0.18 & 602.6
& 0.01 & 436.5
\\
 9 (F) 
& 0.45 & \phantom{0}12.0
& 0.55 & 478.6
& 0.00 & -
\\
10 (Ne) 
& 0.66 & \phantom{00}9.5
& 0.33 & 478.6
& 0.00 & -
\\
11 (Na) 
& 0.67 & \phantom{00}7.9
& 0.32 & 416.9
& 0.00 & -
\\
12 (Mg) 
& 0.60 & \phantom{0}11.5
& 0.37 & 501.2
& 0.03 & 436.5
\\
13 (Al) 
& 0.62 & \phantom{0}11.0
& 0.36 & 416.9
& 0.02 & 436.5
\\
14 (Si) 
& 0.45 & \phantom{0}11.5
& 0.20 & 501.2
& 0.36 & 436.5
\\
15 (P) 
& 0.69 & \phantom{0}10.5
& 0.23 & 501.2
& 0.09 & 436.5
\\
16 (S) 
& 0.54 & \phantom{0}10.5
& 0.19 & 416.9
& 0.27 & 436.5
\\
17 (Cl) 
& 0.73 & \phantom{0}10.0
& 0.22 & 575.4
& 0.05 & 436.5
\\
18 (Ar) 
& 0.56 & \phantom{0}10.5
& 0.19 & 575.4
& 0.25 & 436.5
\\
19 (K) 
& 0.75 & \phantom{0}10.0
& 0.22 & 575.4
& 0.03 & 436.5
\\
20 (Ca) 
& 0.48 & \phantom{0}12.0
& 0.23 & 602.6
& 0.29 & 436.5
\\
21 (Sc) 
& 0.64 & \phantom{0}10.0
& 0.35 & 501.2
& 0.01 & 436.5
\\
22 (Ti) 
& 0.56 & \phantom{0}13.2
& 0.28 & 575.4
& 0.17 & 436.5
\\
23 (V) 
& 0.37 & \phantom{0}12.0
& 0.25 & 602.6
& 0.38 & 436.5
\\
24 (Cr) 
& 0.28 & \phantom{0}13.2
& 0.17 & 575.4
& 0.55 & 436.5
\\
25 (Mn) 
& 0.20 & \phantom{0}12.0
& 0.17 & 575.4
& 0.63 & 436.5
\\
26 (Fe) 
& 0.27 & \phantom{0}13.2
& 0.21 & 524.8
& 0.52 & 436.5
\\
27 (Co) 
& 0.61 & \phantom{0}12.0
& 0.25 & 302.0
& 0.15 & 436.5
\\
28 (Ni) 
& 0.74 & \phantom{0}15.8
& 0.04 & 501.2
& 0.22 & 436.5
\\
29 (Cu) 
& 0.76 & \phantom{0}11.0
& 0.24 & 380.2
& 0.00 & -
\\
30 (Zn) 
& 0.63 & \phantom{0}11.5
& 0.37 & 549.5
& 0.00 & -
\\
31 (Ga) 
& 0.66 & \phantom{00}9.5
& 0.34 & 524.8
& 0.00 & -
\\
32 (Ge) 
& 0.58 & \phantom{00}9.1
& 0.42 & 524.8
& 0.00 & -
\\
33 (As) 
& 0.61 & \phantom{0}10.0
& 0.39 & 549.5
& 0.00 & -
\\
34 (Se) 
& 0.45 & \phantom{00}7.9
& 0.55 & 524.8
& 0.00 & -
\\
35 (Br) 
& 0.50 & \phantom{00}7.9
& 0.50 & 575.4
& 0.00 & -
\\
36 (Kr) 
& 0.31 & \phantom{00}7.9
& 0.69 & 436.5
& 0.00 & -
\\
37 (Rb) 
& 0.34 & \phantom{00}9.1
& 0.66 & 302.0
& 0.00 & -
\\
38 (Sr) 
& 0.11 & \phantom{00}7.9
& 0.89 & 602.6
& 0.00 & -
\\
39 (Y) 
& 0.09 & \phantom{00}7.9
& 0.91 & 602.6
& 0.00 & -
\\
40 (Zr) 
& 0.08 & \phantom{00}7.9
& 0.92 & 602.6
& 0.00 & -
\\
41 (Nb) 
& 0.10 & \phantom{00}8.3
& 0.90 & 758.6
& 0.00 & -
\\
42 (Mo) 
& 0.09 & \phantom{00}7.9
& 0.91 & 602.6
& 0.00 & -
\\
43 (Tc) 
& 0.00 & -
& 1.00 & 316.2
& 0.00 & -
\\
46 (Pd) 
& 0.10 & \phantom{00}7.6
& 0.90 & 660.7
& 0.00 & -
\\
50 (Sn) 
& 0.08 & \phantom{00}7.6
& 0.92 & 660.7
& 0.00 & -
\\
56 (Ba) 
& 0.05 & \phantom{00}7.9
& 0.95 & 758.6
& 0.00 & -
\\
58 (Ce) 
& 0.06 & \phantom{00}7.9
& 0.94 & 758.6
& 0.00 & -
\\

    \enddata
    \tablecomments{Fractional contributions $f$ and 50\% delay times $t_\mathrm{50}$ for selected elements through different nucleosynthetic channels: core-collapse supernovae (CCSN) and massive star winds, AGB stars, and thermonuclear supernovae (TNSN).}
\end{deluxetable}

We then use the same computational setup described in \autoref{ssec:delay_times} to compute delay times for a range of elements of observational interest from atomic number $Z=6-60$. In this calculation we separate out the contributions from core collapse supernovae, thermonuclear supernovae, and AGB stars; for each channel, we identify its fractional contribution to the mass budget for each element (up to a time of 10 Gyr), and we compute the characteristic delay time after star formation for each combination of channel and element, defined as the time after star formation by which that channel has returned 50\% of its eventual total contribution. We do not include neutron star mergers in our model, since the delay time distribution and yields for these are both extraordinarily uncertain, and for this reason we omit for our analysis elements for which the r-process is thought to make a substantial contribution. We summarize the mass fractions and delay times for the elements we include in \autoref{tab:delay_times}. Following \citetalias{Krumholz18a} and \citet{Li25c}, we adopt injection widths $\sigma_i = 60$ pc for core collapse or thermonuclear supernova injection and $\sigma_i = 1$ pc for AGB stars. 

\begin{figure*}
    \includegraphics[width=\textwidth]{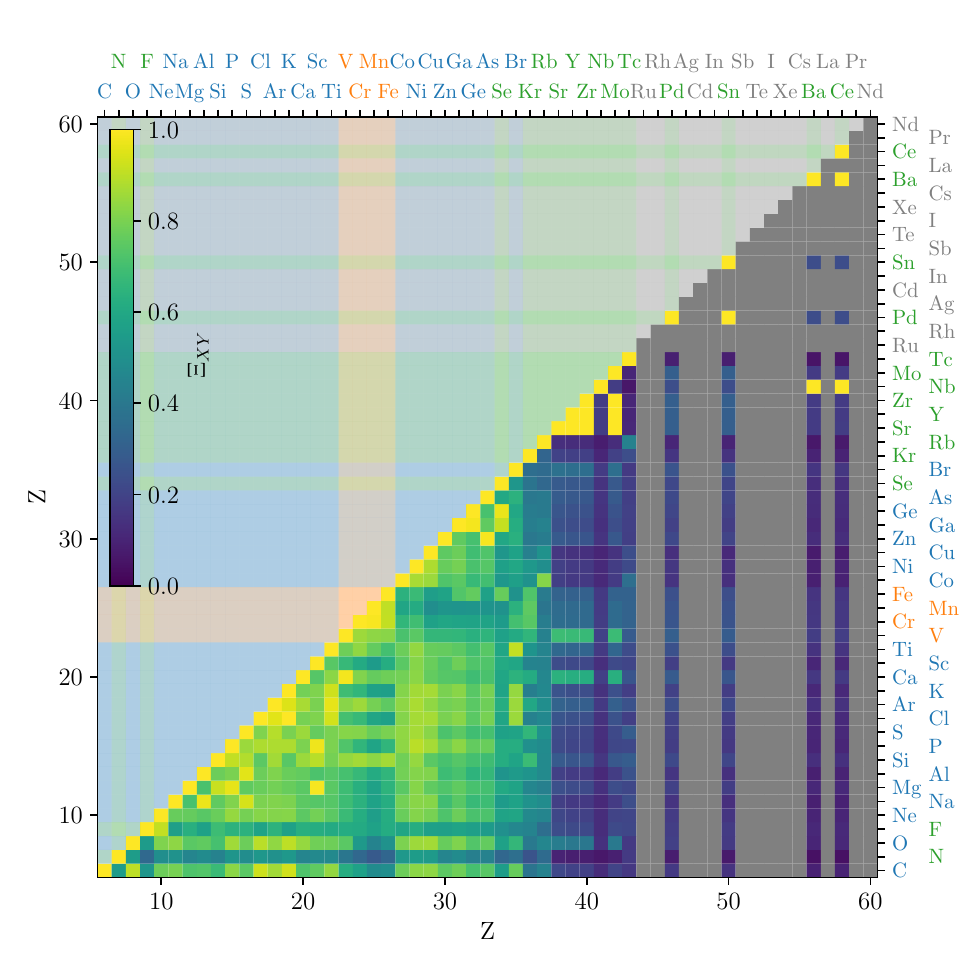}
    \caption{The color map in the lower right corner shows the predicted abundance cross-correlation $\Xi_{XY}$ between elements at zero lag. Elements are ordered by atomic number on the left / bottom axes, and element symbols are shown on the right / top axes; symbols are colored by the dominant source in a Solar-metallicity stellar population: core-collapse supernovae (blue), thermonuclear supernovae (orange), and AGB stars (green). Elements with a significant contribution from the r-process, which we omit from our model, are shown in gray. Shading in the upper left corner of the plot indicates the dominant production channel for each element, and is provided to guide the eye. Note that the values shown here are true cross-correlations, without including the effects of measurement errors.\\
    \label{fig:cross_corr}
    }
\end{figure*}

We plot the cross-correlations we obtain from these procedures in \autoref{fig:cross_corr}. We caution that these are the true cross-correlations, with no contribution from measurement errors; we account for these in \autoref{ssec:obs_eff_gas} and \autoref{ssec:obs_eff_stars}. From the figure, it is clear that elements with a given nucleosynthetic origin are strongly correlated with other elements with the same origin, and significantly less correlated with elements with different nucleosynthetic origins. For example, O (from core-collapse SNe) is $\gtrsim 80\%$ correlated with C, Ne, and S, which have similar origins, but the correlation drops to $\sim 50\%$ for iron-peak elements such as Mn and Fe, which originate in thermonuclear SNe, and to $\lesssim 40\%$ for AGB elements such as Nb, Mo, or Ba. Similarly, V, Cr, Mn, and Fe (thermonuclear-dominated) are $\sim 80\%$ correlated with one another, but only $\sim 50\%$ correlated with core collapse supernova elements such as O, and $\lesssim 40\%$ correlated with AGB elements.

The AGB elements are very well-correlated with other AGB elements that are produced by stars in a similar mass range (e.g., Y and Sr have a correlation near unity), and poorly correlated with anything else; this is partly a result of the small injection width for AGB stars, which amplifies both correlations and anti-correlations depending on differences in the delay time. However, we caution that this effect is likely overestimated in our models, for two reasons. First, we simplify by treating the injection of elements by AGB stars as occurring at a single point in time, rather than with a more realistic time distribution. A more distributed-in-time injection distribution would likely moderate this effect, a point to which we return in \autoref{ssec:caveats}. Second, our AGB delay times for different elements are artificially clustered together due to the relatively coarse mass gridding for the AGB yield models on which we rely \citet{Doherty14a, Karakas16a}. Due to this coarse gridding it may be the case that a single mass model in the grid is responsible for most of the yield of a given element, and if that happens then all elements for which that single mass model is dominant will show nearly-identical delay times. This in turn produces the blockiness visible in \autoref{fig:cross_corr}. This would presumably be reduced by more finely-sampled yield tables, which would allow slightly different delay times. Nonetheless, the qualitative effect is real: elements produced by stars with similar masses and thus similar lifetimes should in general be well-correlated with one another, and more poorly correlated with elements produced by stars that have different lifetimes.

The finding that elements can be divided into groups based on nucleosynthetic origin and that these groups are highly internally correlated is qualitatively consistent with the results of both high-resolution simulations \citep{Zhang25a} and recent analyses of observations \citep[e.g.,][]{Ting22a, Mead25a}. The model presented here provides a clear physical picture for why this should be: the level of correlation is dictated by the combination of time delays between injection and turbulent diffusion in the ISM. Elements with similar nucleosynthetic sources are injected at similar times, and thus there is relatively little time for interstellar turbulence to alter the elemental distributions between successive injection events. The elemental patterns thus reflect the underlying distribution of star formation, and since pairs of elements are ultimately injected by the same stellar populations, they wind up correlated with each other. By contrast, elements that are injected with different delay times wind up only poorly-correlated with one another, because there is enough time between when the first and second elements in a pair are injected for interstellar turbulence to alter the spatial distribution.

From the form of \autoref{eq:crosscorr_zero_dim}, we can even specify a bit more precisely what is meant by ``close together'' in time for this purpose. The difference in injection timescales enters the correlation via the terms containing $|\ell_{X,i}^2 - \ell_{Y,j}^2|$ and $\sigma_v |\Delta t_{X,i} - \Delta t_{Y,j}|$ that appear in \autoref{eq:alpha}-\autoref{eq:gamma}; the former term represents decorrelation due to metals having time to diffuse during the times between when elements X and Y are injected, while the latter represents decorrelation do to the stars moving between the two injection times. Recalling that $\ell_{X,i}^2 = \kappa \Delta t_{X,i}$, and examining the equations, we see that diffusion decorrelation is significant if the characteristic distance over which diffusion transports the elements over the delay time, $\sqrt{\kappa |\Delta t_{X,i} - \Delta t_{Y,j}|}$ is larger than the characteristic size of the injection regions, $\sqrt{\sigma_i^2 + \sigma_j^2}$. Similarly, stellar drift is significant if the characteristic drift distance, $\sigma_v |\Delta t_{X,i} - \Delta t_{Y,j}|$, is larger than both the injection region size and diffusion distance. For our fiducial values of $\kappa$ and $\sigma_v$ above, and for injection regions of size $\sigma_i \approx 60$ pc as appropriate for supernovae, we find that diffusion is important if $|\Delta t_{X,i} - \Delta t_{Y,j}| \gtrsim 10$ Myr, while drift is important if $|\Delta t_{X,i} - \Delta t_{Y,j}| \gtrsim 20$ Myr. This explains why the elements injected primarily by core-collapse supernovae and by thermonuclear supernovae are all well-correlated internally, since the time delays between different elements internal to one of these groups are only $\sim 10$ Myr. However, we can also see that the correlation only decays logarithmically in time at times $\gtrsim 10$ Myr, which is why despite that gap of a few hundred Myr between core-collapse and thermonuclear supernova production, the elements produced are still somewhat correlated.

\subsection{Observational effects in gas}
\label{ssec:obs_eff_gas}

Our predictions thus far do not take into account any observational effects, but these cannot be neglected if we are to carry out meaningful comparisons to observations. We therefore now generalize our predictions to include these effects, building on the approach developed in Appendix D of \citet{Li21a} and Appendix C of \citet{Li25c}. For measurements of metals in gas, there are two distinct sources of error that we must consider: beam smearing and errors in the calibrations used to convert measurable line fluxes to metal abundances.

\subsubsection{Metallicity errors}

For errors in metallicity calibrations, if we make the simple assumption that the errors for different metals are independent\added{\footnote{\added{We caution that, while needed in order to render the problem analytically-tractable, this assumption is fairly strong and is probably not true in detail. For example, if the dominant source of error in gas abundance maps is contamination by diffuse ionised gas, this contamination will likely affect measured elemental abundances in correlated ways, since in the regions subject to contamination abundance inferences for all elements may be similarly affected.}}} \added{and can be approximated as Gaussian}, the effects of errors on element-element correlations are easy to compute. Note that the cross-correlation (\autoref{eq:crosscorr_reduced}\added{)} depends on three terms: $\langle \overline{S_X(\mathbf{r}+\mathbf{r}') S_Y(\mathbf{r}')}\rangle$, $\langle S_X\rangle\langle S_Y\rangle$, and $\sigma_X\sigma_Y$. Under the assumption that errors in elements $X$ and $Y$ are uncorrelated, errors have no effect on the first term since the average will be zero, and \added{any shift in the mean metallicities of elements due to error distributions not being centered exactly at zero will lead to changes in the two terms in the numerator that cancel}. The sole effect of errors then is to increase the dispersions $\sigma_X$ and $\sigma_Y$ of individual elements. Quantitatively, if errors in calibrators increase the measured variance in the abundance of $X$ by a factor $\phi_X$, i.e., if observed and true values of $\sigma_X^2$ are related by $\sigma^2_{X,\mathrm{obs}} = \phi_X \sigma^2_X$, then the effect of metallicity uncertanties is simply to reduce the measured cross-correlation $\Xi_{XY}$ by a factor of $\sqrt{\phi_X \phi_Y}$ relative to the true value.\footnote{Of course if $X=Y$, then $\Xi_{XY}$ remains unity, since in this case the errors are perfectly correlated, and thus $\langle S_X^2\rangle - \langle S_X\rangle^2$ and $\sigma_{S_X}^2$ change by the same factor.} In reality the effect is likely to be weaker than this, because the errors are likely correlated rather than uncorrelated, since they arise for example from errors in the derived electron temperature which are likely to be in the same direction for multiple elements -- see \citet{Li25c} for further discussion.

\subsubsection{Beam smearing}

We idealize the effects of beam smearing as convolution of the true metal field with a Gaussian filter with some specific width $\sigma_\mathrm{beam}$ (which is the Gaussian width, not the full width at half maximum). In Fourier space this convolution is equivalent to multiplication by a Gaussian given by (in our dimensionless unit system) $e^{-s_\mathrm{beam}^2 k^2/2}$, where $s_\mathrm{beam} = \sigma_\mathrm{beam} (\Gamma/\kappa)^{1/4}$. Thus the beam-convolved Fourier-transformed metal field becomes $\tilde{S}_{X,\mathrm{beam}} = \tilde{S}_X e^{-s_\mathrm{beam}^2 k^2/2}$. Examining the form of $\tilde{S}_X$ (\autoref{eq:fourier}), it is clear that this multiplication is fully equivalent to making the substitution $s_j^2 \to s_j^2 +s_\mathrm{beam}^2$, and thus the effects of beam smearing can be incorporated simply by making this substitution in all expressions involving $s_i$ and $s_j$, or equivalently for dimensional variables by replacing $\sigma_i^2$ with $\sigma_i^2 + \sigma_\mathrm{beam}^2$.

Qualitatively, the effect of beam-smearing is to artificially increase the measured cross-correlation between two elements relative to the true one. This increase is significant if $\sigma_\mathrm{beam}^2 \gtrsim \min(\sigma_i^2, \sigma_j^2)$, so accurate measurements of the gas-phase cross-correlation between elements would require very high resolution ($\sim 1$ pc) if one of the elements in question originated in AGB stars, but significantly lower resolution (but still challenging -- $\sim 50$ pc) for elements whose origins are found primarily in supernovae. In practice the only element with a significant AGB contribution that is measurable in gas is nitrogen, and this result means that it will be important to consider the effects of beam smearing in future measurements of N/O correlations. 

\begin{figure}
    \includegraphics[width=\columnwidth]{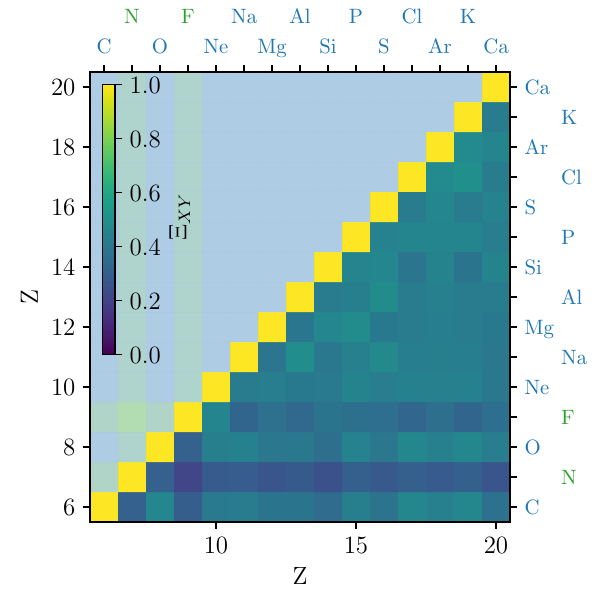}
    \caption{Same as \autoref{fig:cross_corr}, but now focusing only on elements in the range of atomic number that is readily observable in the gas phase, and including the effects of both observational errors characterized by $f=2$ for all elements (see main text) and the effects of smearing by a beam with a FWHM of 20 pc.\\
    \label{fig:cross_corr_gas}
    }
\end{figure}

To illustrate the effects of beam smearing and finite uncertainties in derive abundances, in \autoref{fig:cross_corr_gas} we show sample cross-correlations predicted between elements in the range of $Z$ accessible in gas (from C to Ca), and computed for smearing with a beam of FWHM 20 pc (corresponding to $\sigma_\mathrm{beam} = 10.2$ pc) and using an uncertainty factor $f_X=2$ for all elements. The beam smearing is typical of the resolution achieved by metallicity measurements using MUSE, and the uncertainty is typical of values deduced from analysis of the two-point correlation functions of single elements \citep[e.g.,][]{Li23a, Li25c}. We see that the effect of these two sources of error is to decrease the correlations noticeably compared to their true values, with correlations between two elements both injected by core-collapse SNe dropping from $\Xi_{XY} \gtrsim 80\%$ to $\Xi_{XY} \approx 40-50\%$ and those between this group and AGB elements such as N dropping from $\Xi_{XY}\approx 50\%$ to $\Xi_{XY} \approx 30\%$. Again, however, we caution that our simple assumption of uncorrelated errors likely overestimates the magnitude of the effect, and so in reality the measured correlations may be stronger than suggested in \autoref{fig:cross_corr_gas}.

Interestingly, with a bit of algebra one can show that in the limit of large beam smearing, $s_\mathrm{beam}\to\infty$, the cross-correlation approaches a constant value,\footnote{Note that this is ignoring noise. If there is finite noise, then the limiting value will be lower, with an exact value depending on how the noise level diminishes with $s_\mathrm{beam}$ compared to the cross-correlation signal.}
\begin{equation}
    \lim_{s_\mathrm{beam}\to\infty} \Xi_{XY} = \frac{2\ell_\mathrm{corr}^2 - \ell_X^2-\ell_Y^2 - |\ell_X^2-\ell_Y^2|}{2 \sqrt{(\ell_\mathrm{corr}^2-\ell_X^2)(\ell_\mathrm{corr}^2-\ell_Y^2)}}.
\end{equation}
Note that $\sigma_u$ drops out of this expression, meaning that stellar wander becomes unimportant and the cross-correlation is determined solely by the time delays between element injections and the overall correlation length of the galaxy. This may prove useful in measuring time delays in the future.

\subsection{Observational effects in stars}
\label{ssec:obs_eff_stars}

For observations of stellar abundance correlations, we have two distinct sources of error: one in measuring the abundances themselves, and one in sample selection. 

\subsubsection{Abundance uncertainties}
\label{sssec:stellar_abundance_uncertainties}

Abundance errors are the simpler of the two, and can be handled by much the same formalism as developed in \autoref{ssec:obs_eff_gas} for errors in gas phase distributions, i.e., if measurement uncertainties increase the measured variance in some element $X$ by a factor $\phi_X$ relative to its true value, and uncertainties for different elements $X$ and $Y$ are independent, then the effect of measurement errors is to reduce the correlation $\Xi_{XY}$ by a factor of $\sqrt{\phi_X \phi_Y}$. The only difference for stellar abundances is that the uncertainties are often known, at least approximately, as absolute errors in units of dex, and thus we must deduce values of $\phi_X$ from these uncertainties.

To do so, we make use of the relation derived in \citetalias{Krumholz18a} (their equation 106) between the dimensionless and absolute variances:
\begin{equation}
    \sigma_{\log Z} = \frac{\sigma_{S_X,\mathrm{obs}}}{\langle S_X\rangle \ln 10},
\end{equation}
where $\sigma_{\log Z}$ is the observational error in the logarithmic abundance in units of dex and $\sigma_{S_X,\mathrm{obs}}$ is the scatter in the dimensionless metallicity this level of observational error would induce. We therefore have
\begin{equation}
    \phi_X = 1 + \left(\frac{\langle S_X\rangle \ln 10}{\sigma_{S_X}}\right)^2 \sigma_{\log Z}^2.
    \label{eq:f_from_sigma}
\end{equation}
To evaluate the ratio $\langle S_X\rangle /\sigma_{S_X}$ that appears in this expression, we must adopt values for $\sigma_w^2$, the variance in the mass of stars formed at different star formation events, and $\Gamma$, the number of star formation events per unit area per unit time; the former is required because $\sigma_{S_X}^2 \propto 1 + \sigma_w^2$ (\autoref{eq:sigma_sx_dim}), while the latter is required because the dimensional version of \autoref{eq:SX_avg} is
\begin{equation}
    \langle S_X\rangle = \sqrt{\frac{\Gamma}{\kappa}}\left(\ell_\mathrm{corr}^2 - \sum_{j=1}^{N_\mathrm{ch}} \ell_{X,j}^2\right).
\end{equation}
For this purpose we adopt $\sigma_w^2 = 120$ and $\Gamma = 3$ Myr$^{-1}$ kpc$^{-2}$, which we compute from the formalism of \citetalias[section 3.1.2]{Krumholz18a} assuming that star formation event masses range from $100 - 10^6$ M$_\odot$ with a mass function $dn/dM_*\propto M_*^{-2}$. These values allow us to evaluate $f$ for any given level of observational error $\sigma_{\log Z}$, and thence to compute the correlations we would expect to measure in a survey that achieves a given level of error.

\begin{figure*}
    \includegraphics[width=\textwidth]{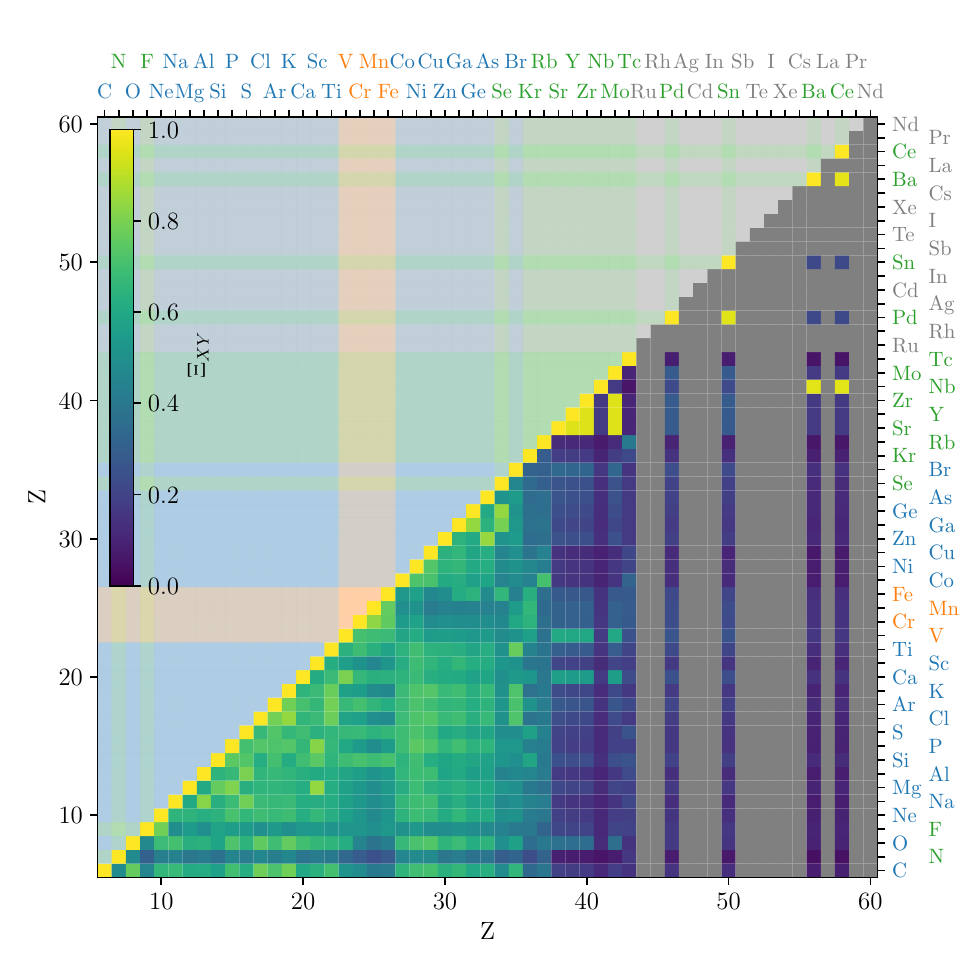}
    \caption{Same as \autoref{fig:cross_corr}, but showing the expected cross-correlation that would be measured in a survey with uniform errors of $\sigma_{\log Z} = 0.01$ dex in all elements.\\
    \label{fig:cross_corr_star_0.01}
    }
\end{figure*}

\begin{figure*}
    \includegraphics[width=\textwidth]{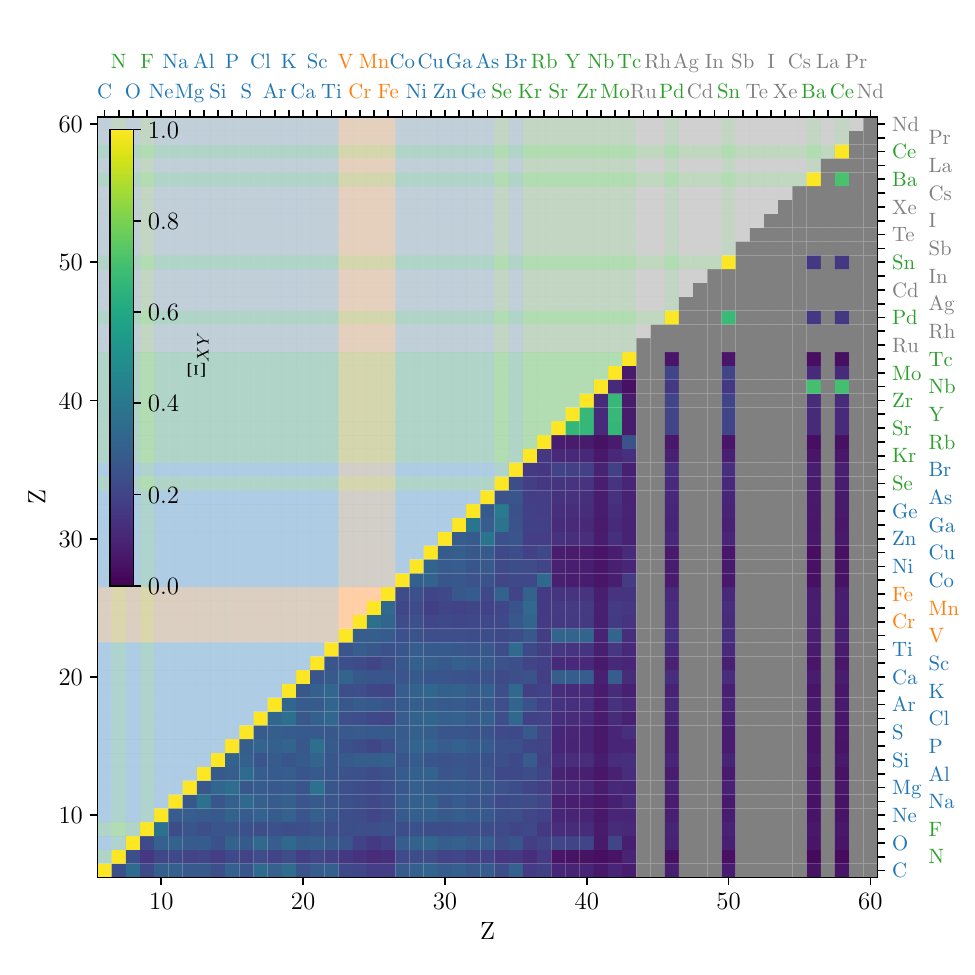}
    \caption{Same as \autoref{fig:cross_corr}, but showing the expected cross-correlation that would be measured in a survey with uniform errors of $\sigma_{\log Z} = 0.03$ dex in all elements.\\
    \label{fig:cross_corr_star_0.03}
    }
\end{figure*}

In \autoref{fig:cross_corr_star_0.01} and \autoref{fig:cross_corr_star_0.03} we show our estimates of the measured element-element cross-correlation in surveys with errors of $\sigma_{\log Z} = 0.01$ and 0.03 dex for all elements, respectively. Comparing these two figures to \autoref{fig:cross_corr}, which shows the true cross-correlation, we see that an error of 0.01 dex has relatively minimal effects on the cross-correlations, reducing them by $\sim 10-20\%$ on average. An error of 0.03 dex is much more serious, and suppresses almost all cross-correlations to below 50\%. These cross-correlations would still be detectable in a sufficiently large sample, and one could still for example measure the greater cross-correlation within nucleosynthetic groups compared to across them, but the absolute values of the correlations are substantially reduced.

\added{For example, our model predicts that the true correlation of C and O (two elements from the same nucleosynthetic channel) is 0.90, but uncertainties of 0.01 and 0.03 dex would reduce this to 0.76 and 0.34, respectively. By contrast, for C and N (from two different nucleosynthetic channels) our predicted true cross-correlation is 0.55. This only changes slightly to 0.48 with a 0.01 dex uncertainty, but it falls to 0.24 with 0.03 dex uncertainty -- not so different from the C-O correlation of 0.34 at the same uncertainty level. Thus it should be relatively straightforward to detect the difference in typical size between intra- and inter-group correlations with 0.01 dex uncertainty, but this difference is significantly harder to measure if the uncertainty is 0.03 dex.}

As in \autoref{ssec:obs_eff_gas} we caution that these results are computed under the assumption that the errors in different elements are uncorrelated, which is unlikely to be true in detail. Depending on how the errors are correlated, the effects could be either smaller or larger than what we have computed, and the size and direction of the effect may be different for different element pairs. Moreover, these results are relatively sensitive to the choice of $\sigma_w^2$, which controls the intrinsic spread in the stars. When this value is large compared to unity (which is almost certainly the case), the value of $f$ and thus the amount by which a given level of observed error reduces the measured correlation scales inversely with $\sigma_w^2$. Thus while we have provided a theoretical calculation, in an observational survey a better strategy is likely to be to estimate the intrinsic scatter in the sample directly from the data, and use this in turn to estimate the effects of observational error on the measured correlations.

\subsubsection{Age spreads}
\label{sssec:stellar_age_spreads}

Uncertainties arising from sample selection for stars are significantly more challenging to handle. Our calculations of elemental abundance correlations are instantaneous in time, and thus in principle we should be comparing to mono-age stellar populations. In practice, however, we cannot in most cases measure stellar ages to very high precision, and any heterogeneity in age in a sample introduces the risk that measured correlations will mostly reflect the effect that all elements are correlated with one another simply because they are correlated with the galactic star formation history. For example even though we expect a very small correlation between the spatial distributions of, for example, O (CCSN-dominated) and Ce (AGB-dominated) in the ISM or a mono-age stellar population, in a sample that includes both old, metal-poor stars and young, metal-rich ones we may nevertheless observe a strong correlation between these two elements simply because the sample includes both stars formed early in the Galaxy's history when there had been little time for either supernova or AGB enrichment, while some formed much more recently and thus have had longer for both enrichment channels to operate. Correlations in an age-heterogenous sample of stars are sensitive both to the spatial distribution of element abundances and to the overall star formation and enrichment history of the Galaxy, and it is far from easy to disentangle these two contributors.

In principle we could attempt to fold this evolution into our model and directly predict the correlations in age-heterogenous populations, since the model predicts the distribution of element abundances at all times -- indeed, in \citetalias{Krumholz18a} we use this formalism to show that the correlation time of elemental abundance distributions is $\sim 200$ Myr. However, to expect to reproduce observations of metal correlations in a realistically age-heterogenous sample, we would need our model to be reliable over multi-Gyr timescales. This seems improbable, since many of the simplifying assumptions we make -- for example ignoring heavy element loss via winds, dilution by cosmological accretion, and constant star formation rate -- may be reasonable over $\sim 200$ Myr but are clearly not reasonable over $\sim 5$ Gyr. For example, our assumption of constant star formation rate with no metal loss or dilution implies that the mean abundances of all elements and the squared correlation length both increase linearly with time, but observations indicate that Milky Way-mass galaxies have had nearly constant mean oxygen abundance since $z\sim 1$ \citep[e.g.,][]{Zahid11a, Zahid13a}, while analyses of cosmological simulations similarly show that correlation lengths are also nearly constant over this epoch \citep{Li24a}, consistent with a view that over cosmological timescales the balance between metal production and other processes leads to a quasi-steady metallicity distribution \citep[e.g.,][]{Sharda21a, Johnson25a}. 

Given the challenges of making an analytic model that can handle heterogenous-age samples, a better strategy for comparing models to observations is to try to obtain as close to mono-age observational sample as possible. There are various strategies available for this purpose, from measuring abundances in B stars that are necessarily young \citep[e.g.,][]{Nieva12a, Lyubimkov13a, Takeda16a, Wessmayer22a, Liu22a, Elmasli23a}. However, at present only a limited range of elements are accessible in such stars, and even for them accuracy is considerably lower than can be achieved for cooler stars. The challenges and systematics of homogenizing the ages in such cooler stellar samples is beyond the scope of this paper, but is likely to be a substantial source of error in comparing to our model.

\subsection{Comparison to measurements}
\label{ssec:obs_comparison}

As our discussion of observational effects in the preceding two sections makes clear, a quantitative comparison between our model and observations requires careful attention to observational errors, since measuring abundances with enough fidelity to recover the true level of correlation (as shown in \autoref{fig:cross_corr}) is extremely challenging with current observational capabilities. A self-consistent approach requires fitting observations to one of the error-convolved model predictions developed in the previous section, leaving the (usually unknown) level of error as a nuisance parameter in the fit, following for example \citet{Li21a, Li23a, Li25c}. Nonetheless it is useful to ask if our predictions are at least qualitatively consistent with observations.

\subsubsection{Gas observations}
\label{sssec:gas_observations}

In gas, \citet{Li25c}\footnote{\citet{Bresolin25a} also measured maps of N and O from which a cross-correlation could be computed. However, they do not report the cross-correlation in their data, only a comparison between the correlation lengths of N and O derived from the two-point correlation functions.}  have measured the cross-correlations of N, O, and S. Prior to correcting for observational uncertainties, they measure $\Xi_\mathrm{OS} = 0.88$, $\Xi_\mathrm{NO} = 0.31$, and $\Xi_\mathrm{NS} = 0.21$, while with uncertainty corrections derived assuming that the errors on all elements are independent they have $\Xi_\mathrm{OS} = 1.9$, $\Xi_\mathrm{NO} = 0.69$, and $\Xi_\mathrm{NS} = 0.45$ for the correlations.\footnote{Astute readers will note that their value of $\Xi_\mathrm{OS}$ is mathematically forbidden, since cross-correlations cannot exceed unity. In their paper \citet{Li25c} explain that this value is a result the uncertainty correction ``overshooting'', because the assumption that the errors in the different elements are completely independent is certainly incorrect.}
whereas our predicted true correlations for these elements and our fiducial parameters are 0.77, 0.54, and 0.49. Thus our model recovers the important qualitative result that O and S are very highly-correlated, while the correlations are lower but still non-negligible for N versus O and N versus S. The exact correlation values do not match, which is not surprising given both the simplicity of our model and the complexities of deriving elemental abundances from line fluxes, but the qualitative agreement is highly encouraging.

\subsubsection{Stellar observations}

For stellar correlations, we compare our predicted cross-correlations to observations drawn from two data sets. The first comes from \citet{Mead25a}, who measure element-element correlations in a sample of Solar twins drawn from \citet{Bedell18a}. After removing elements with significant r-process contributions, the sample provides cross-correlations between the abundances of 22 elements: 14 dominated by core-collapse supernovae and massive stars (C,\footnote{\citeauthor{Mead25a}~provide two estimators for the abundance of C, based on C~\textsc{i} and CH lines; we use the latter due to its smaller statistical uncertainty.} O, Na, Mg, Al, Si, S, Ca, Sc, Ti, Co, Ni, Cu, and Zn), three dominated by thermonuclear supernovae (V, Cr, and Mn), and five dominated by AGB stars (Sr, Y, Zr, Ba, and Ce). 

Our second comparison data set comes from the fits of \citet{Ting22a}, who construct probabilistic models through normalizing flows to represent abundance correlations in the APOGEE sample, covering the 13 elements O, Na, Mg, Al, Si, S, Ca, V, Cr, Fe, Co, Ni, and Cu; of these, four -- V, Fe, Cr, and Mn -- are thermonuclear supernova-dominated, while the remaining nine are core collapse-dominated. While this data set is heterogenous in metallicity and other stellar parameters, the model makes it possible to extract the correlations expected at a set of fixed stellar parameters. To do so, we first draw samples from the distribution constructed in \citeauthor{Ting22a} to represent the APOGEE data restricted to the range of stellar parameters $T_\mathrm{eff} = 4100 - 4300$ K, $\log (g/\mathrm{cm}\,\mathrm{s}^{-2}) = 1.5 - 1.7$, and $\left|[\mathrm{Fe/H}]\right| < 0.1$ dex or 0.05 dex, where $[\mathrm{X/H}]$ has its usual meaning of the logarithmic abundance relative to Solar. The choice of window size in [Fe/H] reflects a balance between the competing imperatives of making the sample as homogenous as possible to eliminate time evolution effects (favoring smaller windows) without masking true flucutations that contribute to correlations (favoring larger windows); we consider two values of the window in [Fe/H] to explore how different choices in balancing these factors affects the results. We then compute our correlations from samples drawn from the central 5th to 95th percentile of the abundances of all remaining elemental abundances.

To generate predictions against which we can compare these data, we first compute the intrinsic correlation for each element pair in our model, using the same parameters as described in \autoref{ssec:crosscorr_table} and to generate \autoref{fig:cross_corr}. We then correct our predicted true correlations $\Xi_{XY,\mathrm{true}}$ for the effects of measurement uncertainties in the abundance each element for each observational sample, following the procedure described in \autoref{sssec:stellar_abundance_uncertainties}, i.e., we convert the logarithmic uncertainty for each element $\sigma_{\log Z}$ to an uncertainty factor $\phi_X$ using \autoref{eq:f_from_sigma}, and then divide our predicted cross-correlation for each element pair by a factor of $\sqrt{\phi_X \phi_Y}$, so that our final predicted cross-correlation is $\Xi_{XY,\mathrm{pred}} = \Xi_{XY,\mathrm{true}} / \sqrt{\phi_X \phi_Y}$. We report the element-by-element logarithmic abundance uncertainties that enter into this calculation in \autoref{tab:stellar_uncertainties}; for the \citet{Mead25a} data set we use the median uncertainties from \citet{Bedell18a}, while for the \citet{Ting22a} sample we use the statistical uncertainties from \citet{Jonsson20a}. 

\begin{deluxetable}{lcc}
    \tablecaption{Statistical uncertainties on elemental abundances for stellar samples
    \label{tab:stellar_uncertainties}
    }
    \tablehead{
    & \colhead{\citet{Mead25a}} 
    & \colhead{\citet{Ting22a}} \\
    \colhead{Z} &
    \colhead{$\sigma_{\log Z}$} &
    \colhead{$\sigma_{\log Z}$}
    \\
    & \colhead{[$10^{-2}$ dex]}
    & \colhead{[$10^{-2}$ dex]}
    }
    \startdata
    6 (C) & 1.000 & - \\
8 (O) & 1.300 & 0.958 \\
11 (Na) & 1.400 & 3.600 \\
12 (Mg) & 1.200 & 1.070 \\
13 (Al) & 1.000 & 1.800 \\
14 (Si) & 0.600 & 1.430 \\
16 (S) & 1.800 & 3.100 \\
19 (K) & - & 2.980 \\
20 (Ca) & 0.800 & 1.140 \\
21 (Sc) & 1.500 & - \\
22 (Ti) & 0.800 & - \\
23 (V) & 0.900 & 3.730 \\
24 (Cr) & 0.800 & 2.950 \\
25 (Mn) & 0.800 & 1.260 \\
26 (Fe) & 0.300 & 0.800 \\
27 (Co) & 0.850 & 3.080 \\
28 (Ni) & 0.700 & 0.972 \\
29 (Cu) & 1.500 & 3.010 \\
30 (Zn) & 1.400 & - \\
38 (Sr) & 0.800 & - \\
39 (Y) & 1.100 & - \\
40 (Zr) & 1.400 & - \\
56 (Ba) & 1.100 & - \\
58 (Ce) & 1.800 & - \\

    \enddata
    \tablecomments{Values adopted from \citet{Bedell18a} for the \citet{Mead25a} sample, and from \citet{Jonsson20a} for the \citet{Ting22a} sample.\\}
\end{deluxetable}

\begin{figure}
    \includegraphics[width=\columnwidth]{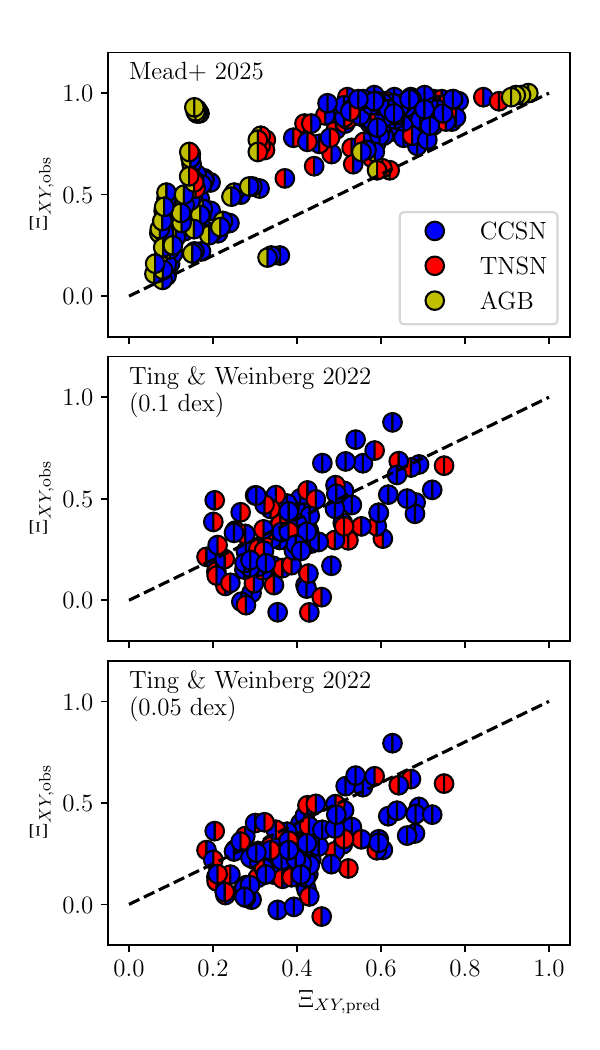}
    \caption{Comparison between the elemental cross-correlations predicted by our model ($\Xi_{XY,\mathrm{pred}}$, horizontal axis) and observed estimates drawn from three samples ($\Xi_{XY,\mathrm{obs}}$, vertical axis). Each panel shows a different observational sample; see main text. Points are colored by the predominant origins of the two elements being correlated -- see main text for a full list of the elements included. The dashed black line indicates the 1-to-1 relation. Auto-correlations between an element and itself, which would all appear at exactly coordinate $(1,1)$, are not included in the plot, so the points near $(1,1)$ visible in the figure consist of different elements for which both our model prediction and the measurements yield very high correlations. \\
    \label{fig:Xi_comp}
    }
\end{figure}

We plot the results in \autoref{fig:Xi_comp}, which shows qualitatively reasonable agreement\added{. The major trend visible in all three sets of observations is that pairs of elements from the same nucleosynthetic group (matching-color circles in the Figure) are more correlated (and thus lie at higher in the plot) than element pairs from different groups (different-color circles).} The effect is particularly clear in the \citeauthor{Mead25a} sample, which includes AGB elements that correlate poorly with CCSN elements. Quantitatively, our model does a good job of \added{reproducing this trend: our predicted correlations are also larger (and thus lie further right in the plot) for same-channel element pairs than for different-channel pairs. In addition to capturing this major qualitative feature, our model also make reasonable predictions for the correlations pair-by-pair}; the Pearson correlation coefficients between our predicted cross-correlations and the measured ones are 0.82, 0.58, and 0.60 for the data sets shown in the top to bottom panels of \autoref{fig:Xi_comp}, respectively.

However, it is also clear that there are systematic differences between both the model and the data and between the different data sets, and that details of how we process the data matter. For example the correlations measured in \citeauthor{Mead25a}~are systematically larger than those predicted by the model, while those for \citeauthor{Ting22a} selected using a 0.1 dex window in [Fe/H] scatter evenly about the one-to-one line, and observed correlations derived using a 0.05 dex window are smaller than predicted. While differences between the \citeauthor{Mead25a}~and \citeauthor{Ting22a} data sets could be due to errors in the uncertainty estimates in \autoref{tab:stellar_uncertainties}, or to unaccounted-for correlations between the errors for different elements, such effects cannot explain the offset between the \citeauthor{Ting22a} results obtained using different windows in [Fe/H]. Thus the fact that the two bottom panels in \autoref{fig:Xi_comp} are systematically different suggests that the dominant driver of these differences, and by extension between the data and the model, is almost certainly the challenges of sample selection outlined in \autoref{sssec:stellar_age_spreads}.

The \citeauthor{Mead25a}~sample is selected to be Solar twins and the stars that comprise it are mostly from the Solar neighborhood, which should somewhat limit the age range, but the sample is by no means truly mono-age. Similarly, we have homogenized the \citeauthor{Ting22a} sample by metallicity, effective temperature, and $\log g$, but this in no way guarantees that the sample is mono-age, nor that the range of ages we are effectively selecting matches that of the \citeauthor{Mead25a}~sample. Thus the systematic differences in measured correlations may simply reflect the differing degrees of contamination by the age spread of the stellar population, with the differences between the results derived for \citeauthor{Ting22a} using 0.05 and 0.1 dex windows reflecting different balances between the risk of excluding real intrinsic stellar yield fluctuations versus finding correlations that are driven by chemical evolution signals rather than actual element abundance correlations in the ISM.

For the \citeauthor{Mead25a} data set, we can also identify particular discrepancies between the predicted and observed correlations for AGB elements. For example, the cluster of points in the top panel of \autoref{fig:Xi_comp} for which the predicted correlation is $\approx 20\%$ while \citeauthor{Mead25a} find $\approx 90\%$ consists of the cross-correlations of Y and Zr versus Ba and Ce, while the cluster of points at $\Xi_{XY,\mathrm{pred}}\approx 0.35$ and $\Xi_{XY,\mathrm{obs}}\approx 0.8$ is made up of Cr versus Y and Sr and similar cross-correlations of AGB-dominated with thermonuclear-dominated elements. This may reflect the limitations of our approximation of AGB injection as instantaneous or of star formation as Poissonian in space (see \autoref{ssec:caveats} for further discussion), but may again reflect the effects of contamination by the intrinsic age spread in the sample.

Despite these differences in detail, the overall conclusion we reach from this comparison exercise is that our model successfully reproduces the major qualitative features of the data. These are that there are very strong correlations between elements of the same nucleosynthetic group, moderately strong correlations between core-collapse supernova- and thermonuclear supernova-group elements, and weaker correlations between AGB elements and elements from non-AGB sources.

\subsection{Limitations and future directions}
\label{ssec:caveats}

Our model for element cross-correlations has the virtues of simplicity and exact solubility, but this necessarily entails some compromises that it is important to mention. \added{We provided a full list of what we see see as the major omissions and simplifications in \autoref{ssec:model_basics}, but here we explore the first two of these in more detail and suggest possible paths to remedying them, along with a third issue that is both a limitation and an opportunity.}

A first limitation is that our assumption that star formation events are Poisson distributed is clearly not correct in detail. Star formation in a Milky Way-sized galaxy is concentrated in spiral arms, and thus the locations of star formation events are not Poissonian in space. Both observations \citep[e.g.,][]{Sanchez-Menguiano16a, Ho17a, Ho18a, Wenger19a} and simulations \citep{Yang12a, Orr23a, Zhang25a} suggest that spiral arms leave measurable imprints in galaxy metal fields, an effect that our simple model ignores. In terms the auto- and cross-correlation functions of metal fields, the primary effect of spiral structure is to introduce ``ripples'' in the correlation function at scales that correspond to typical inter-arm spacings, but for the correlation at zero lag -- our primary concern here -- the effect is likely to be an enhancement in elemental cross-correlations relative to our simple model. Such an enhancement will occurs simply because in our simple model where star formation is Poissonian different star formation events are uncorrelated with each other, and thus make zero contribution to $\Xi_{XY}$, but if star formation events are spatially correlated because they are all occurring in the same spiral structures, this will represent an additional contribution to $\Xi_{XY}$. In future work we intend to model this effect by replacing our expectation value for the $\langle e^{i k\cdot(\mathbf{r}_i-\mathbf{r}_{i'})}\rangle$ term in \autoref{eq:crosscorr_gen_1}, which is currently derived under the assumption that $\mathbf{r}_i$ and $\mathbf{r}_{i'}$ are both drawn from a uniform distribution, with a more realistic model.

A second limitation to which we wish to draw reader attention is our treatment of metal injection as occurring at discrete points in time. While this has the advantage of computational simplicity, it is clearly a coarse approximation, particularly for AGB stars whose yields may be spread over multi-Gyr timescales. Fortunately our formalism allows a natural generalization to a continuous distribution: since at no point have we assumed that the number of injection channels $N_\mathrm{ch}$ has any particular value, we are free to take the limit as the number of channels goes to infinity, with each channel injecting an infinitesimal amount of mass. In terms of our mathematical formalism, this amounts to replacing all our terms of the form
\begin{equation}
    \sum_{i=1}^{N_\mathrm{ch}} \sum_{j=1}^{N_\mathrm{ch}} f_{X,i} f_{Y,j} q (\Delta t_{X,i}, \Delta t_{Y,j}),
\end{equation}
appearing for example in \autoref{eq:crosscorr_dim} and \autoref{eq:crosscorr_zero_dim}, with integrals of the form
\begin{equation}
    \iint \frac{df_X}{dt_X} \frac{df_Y}{dt_Y} q(t_X, t_Y) \, dt_X \, dt_Y,
\end{equation}
where $df_X/dt_X$ is the differential increase in mass fraction returned over the time interval from $t_X$ to $t_X + dt_X$, and similarly for $df_Y/dt_Y$. The resulting integrals are analytically intractable, but are straightforward to evaluate numerically given a tabulated mass return over time $df_X/dt_X$. 

Our third and final limitation is with regard to r-process elements. It is clear from observations of the electromagnetic counterpart to GW170817 that r-process elements are likely produced predominantly via the mergers of neutron stars \citep{Kasen17a}, but since we have thus far observed only one such event, it is not possible at this point to make meaningful estimate for the delay time distribution between star formation and element production for this channel. For this reason, we have omitted them from our model here.

However, this represents an opportunity as well as a challenge: we have shown that the degree of correlation between elements is determined by the relative contributions of different nucleosynthetic channels with differing delay times. This means that measurements of that correlation can be used to infer the currently-unknown r-process contribution and delay time for elements with significant a significant r-process component. That is, given a set of measured element-element correlations for elements that include r-process contributions, we can ask what delay time the r-process elements are required to have, and what fractional contribution to each element's abundance they must represent, in order to match the measured level of correlation. Thus stellar samples suitable for measuring element-element correlation \citep[e.g.,][]{Ting22a, Mead25a} can also be used to infer the typical neutron star merger delay. We intend to pursue this problem in future work.

\section{Conclusions}
\label{sec:conclusion}

Observations have begun to explore the structure of elemental abundance space, as characterized by the correlations between the abundances of different elements in both the interstellar medium and in stars. However, there has been little theoretical exploration to date aiming to understand the physical origins of these correlations. In this paper we extend the simple stochastically-forced diffusion model of \citet{Krumholz18a} for metallicity fluctuations in the interstellar medium to the case of multiple elements, and use this model to calculate expected strengths of correlation between different elements.

In our model, elements are correlated because they are ultimately produced by the same stellar populations, which are born at discrete points in space and time. Once formed, those stars begin to release newly-synthesized elements on a range of timescales depending on the nucleosynthetic pathway from which they come, but even the longest nucleosynthetic pathways lead to most newly-formed mass being returned in well under a Gyr, and for many elements considerably less time than that. As a result, stars have not fully randomized their positions between when they release one element and another, and this similar in the positions at which stars release elements ultimately drives the correlation between them. The more similar the nucleosynthetic pathways -- and thus more similar the delay times between two elements -- the more correlated they will be. As a result, elements produced primarily by the same nucleosynthetic pathway wind up highly correlated with one another, and moderately to poorly correlated with elements forged by different nucleosynthetic paths.

We show that the structure predicted by this model agrees reasonably well with recent observations of elemental abundance cross-correlations in both gas and stars. On this basis, we argue that it should be possible to use measurements of the cross-correlation to obtain at least rough constraints on the contributions to the abundances of different elements and the associated delay times for each nucleosynthetic pathway. This is likely to be a particularly powerful technique for elements with r-process origins, for which the astrophysical origin site is now known to be neutron star mergers, but where the yields and the delay time distribution are both highly uncertain. We will explore this constraint further in forthcoming work.

\section*{Data and software availability statement}

The software and source data used to produce all the figures and quantitative analysis in this paper are available on GitHub at \url{https://github.com/markkrumholz/z_fluctuation} under an open source license. This software makes use of the \textsc{NumPy} \citep{Numpy20a}, \textsc{SciPy} \citep{SciPy20a}, \textsc{AstroPy} \citep{Astropy-Collaboration13a, Astropy-Collaboration18a, Astropy-Collaboration22a}, \textsc{slug} \citep{da-Silva12a, Krumholz15b}, and \textsc{Matplotlib} \citep{Hunter07a} packages.

\section*{Acknowledgements}

\added{We thank the referees for helpful comments.} MRK acknowledges support from the Australian Research Council through Laureate Fellowship FL220100020. ZL acknowledges the Science and Technology Facilities Council (STFC) consolidated grant ST/X001075/1.

\appendix

\section{Evaluation of the cross-correlation power spectrum}
\label{app:crosscorr_eval}

In this appendix we step through evaluation of the cross-correlation power spectrum, \autoref{eq:crosscorr_spec}. Inserting our expression for the Fourier transform of the metal field, \autoref{eq:fourier}, in this expression gives
\begin{eqnarray}
    \Psi_{XY}(k) & = & \lim_{R\to\infty} \frac{e^{-2\tau k^2}}{4\pi^3 R^2}
    \left\langle
    \sum_{i=1}^{N_\mathrm{sf}}
    \sum_{i'=1}^{N_\mathrm{sf}}
    w_i w_{i'} e^{(\tau_i + \tau_{i'})k^2 + i \mathbf{k}\cdot(\mathbf{r}_i-\mathbf{r}_{i'})}
    \right.
    \nonumber \\
    & &
    \quad
    \left.
    \sum_{j=1}^{N_{\mathrm{ch}}} \sum_{j'=1}^{N_{\mathrm{ch}}}
    f_{X,j} f_{Y,j'} e^{(\Delta \tau_{X,j} + \Delta\tau_{Y,j'} - s_{j}^2/2 - s_{j'}^2/2)k^2}
    e^{i\mathbf{k}\cdot(\mathbf{u}_i\Delta\tau_{X,j}-\mathbf{u}_{i'}\Delta\tau_{Y,j'})}
    H(\theta_{X,ij}) H(\theta_{Y,i'j'})
    \right\rangle.
    \label{eq:crosscorr_gen_1}
\end{eqnarray}
Because the various random numbers are drawn independently, we can break up the expression above into multiplicative factors that depend only on $N_\mathrm{sf}$, only on $w$, only on $\mathbf{r}$, only on $\mathbf{u}$, and only on $\tau$, and evaluate the averages over these distributions independently. These terms are:
\begin{eqnarray}
    \left\langle N_\mathrm{sf}\right\rangle & = & \pi R^2\tau 
    \label{eq:nsf_exp}
    \\
    \left\langle N_\mathrm{sf}^2\right\rangle & = & \pi R^2\tau(\pi R^2\tau+1) \\
    \left\langle w_i w_{i'} \right\rangle & = & 1 + \sigma_w^2 \delta_{ii'} \\
    \left\langle e^{i\mathbf{k}\cdot(\mathbf{r}_i-\mathbf{r}_i')}\right\rangle & = & \delta_{ii'} + \left[2\frac{J_1(kR)}{kR}\right]^2 \left(1-\delta_{ii'}\right),
    \label{eq:space_avg_term}
    \\
    \left\langle e^{i\mathbf{k}\cdot(\mathbf{u}_i\Delta\tau_{X,j} - \mathbf{u}_{i'}\Delta\tau_{Y,j'})}\right\rangle & = &
    \left[
    1 - 2 k R_{XY,jj'} F(k R_{XY,jj'})
    \right] \delta_{ii'} + e^{-\sigma_u^2 (\Delta\tau_{X,j}^2+\Delta\tau_{Y,j'}^2)k^2/2} (1-\delta_{ii'}),
\end{eqnarray}
where $R_{XY,jj'} \equiv \sigma_u |\Delta\tau_{X,j} - \Delta\tau_{Y,j'}|/\sqrt{2}$ and $F(x)$ is the Dawson integral defined by $F(x)=\int_0^x e^{y^2-x^2}\, dy$, and finally
\begin{eqnarray}
    \lefteqn{
    \left\langle
    e^{(\tau_i + \tau_{i'})k^2}
    \sum_{j=1}^{N_{\mathrm{ch}}} \sum_{j'=1}^{N_{\mathrm{ch}}}
    f_{X,j} f_{Y,j'} e^{(\Delta \tau_{X,j} + \Delta\tau_{Y,j'} - s_{j}^2/2 - s_{j'}^2/2)k^2}
    H(\theta_{X,ij}) H(\theta_{Y,i'j'})
    \right\rangle =
    }
    \nonumber \\
    & & 
    \frac{1}{\tau k^2}
    \sum_{j=1}^{N_{\mathrm{ch}}} \sum_{j'=1}^{N_{\mathrm{ch}}} f_{X,j} f_{Y,j'}
    e^{-(s_{j}^2+s_{j'}^2)k^2/2}
    \left[
    \frac{1}{\tau k^2}
    \left(e^{\Delta\tau_{X,j} k^2} - e^{\tau k^2}\right)
    \left(e^{\Delta\tau_{Y,j'} k^2} - e^{\tau k^2}\right)
    (1 - \delta_{ii'})
    \right.
    \nonumber \\
    & &
    \qquad
    \left.
    {}+
    \frac{1}{2}
    \left(
    e^{(2\tau-|\Delta \tau_{X,j}-\Delta\tau_{Y,j'}|)k^2}
    - 
    e^{(\Delta\tau_{X,j}+\Delta\tau_{Y,j'})k^2}
    \right)
    \delta_{ii'}
    \right].
\end{eqnarray}
Inserting these factors into the cross-correlation power spectrum gives
\begin{eqnarray}
    \Psi_{XY}(k) & = & \frac{e^{-2\tau k^2}}{8\pi^2 k^2}
    \sum_{j=1}^{N_\mathrm{ch}} 
    \sum_{j'=1}^{N_\mathrm{ch}}
    f_{X,j} f_{Y,j'} e^{-(s_{j}^2+s_{j'}^2)k^2/2}
    \nonumber \\
    & &
    \quad
    \left\{
    \vphantom{\frac{8\pi}{k^4}}
    \left(1+\sigma_w^2\right)
    \left(
    e^{(2\tau-|\Delta \tau_{X,j}-\Delta\tau_{Y,j'}|)k^2}
    - 
    e^{(\Delta\tau_{X,j}+\Delta\tau_{Y,j'})k^2}
    \right)
    \left[
    1 - 2 k R_{XY,jj'} F(k R_{XY,jj'})
    \right]
    \right.
    \nonumber \\
    & &
    \qquad
    \left.
    {} +
    \frac{8\pi}{k^4} 
    \left(e^{\Delta\tau_{X,j} k^2} - e^{\tau k^2}\right)
    \left(e^{\Delta\tau_{Y,j'} k^2} - e^{\tau k^2}\right)
    e^{-\sigma_u^2 (\Delta\tau_{X,j}^2+\Delta\tau_{Y,j'}^2)k^2/2} 
    \lim_{R\to\infty} J_1(kR)^2
    \right\}.
\end{eqnarray}
We next insert this power spectrum into the cross-correlation theorem (\autoref{eq:crosscorr_thm}) to compute the cross-correlation function. To evaluate the resulting integral, we note that in the limit $R\to\infty$, the Bessel function $J_1(kR)$ vanishes outside an infinitesimal neighborhood of $k=0$, and we can therefore evaluate the integral involving this term by Taylor expanding the terms by which it is multiplied about $k=0$, which renders the integral analytically tractable. Using this approach, we finally arrive at
\begin{eqnarray}
    \xi_{XY}(r) & = &
    \frac{1+\sigma_w^2}{4\pi \sigma_{S_X}\sigma_{S_Y}}
    \sum_{i=1}^{N_\mathrm{ch}} 
    \sum_{j=1}^{N_\mathrm{ch}}
    f_{X,i} f_{Y,j} 
    \int_0^\infty \frac{J_0(kr)}{k}
    e^{-(s_{X,i}^2+s_{Y,j}^2)k^2/2}
    \nonumber \\
    & & 
    \qquad
    \left(e^{-|\Delta\tau_{X,i}-\Delta\tau_{Y,j}|k^2}
    - e^{(\Delta\tau_{X,i}+\Delta\tau_{Y,j}-2\tau)k^2}
    \right)
    \left[
    1 - 2 k R_{XY,ij} F(k R_{XY,ij})
    \right]
    \,
    dk,
    \label{eq:crosscorr_final}
\end{eqnarray}
At zero lag, $r=0$, this evaluates to
\begin{equation}
    \Xi_{XY} = \frac{1+\sigma_w^2}{8\pi\sigma_{S_X}\sigma_{S_Y}}
    \sum_{i=1}^{N_{\mathrm{ch}}}
    \sum_{j=1}^{N_{\mathrm{ch}}}
    f_{X,i} f_{Y,j}
    \left(
    \ln \alpha_{XY,ij}
    +
    2 \sqrt{\frac{\beta_{XY,ij}^2}{1+\beta_{XY,ij}^2}}\sinh^{-1}\beta_{XY,ij}
    -
    2 \sqrt{\frac{\gamma_{XY,ij}^2}{1+\gamma_{XY,ij}^2}}\sinh^{-1}\gamma_{XY,ij}
    \right)
    \label{eq:crosscorr_zero_final}
\end{equation}
where
\begin{eqnarray}
    \alpha_{XY,ij} & = & \frac{4\tau-2\Delta\tau_{X,i}-2\Delta\tau_{Y,j} + s_{i}^2 + s_{j}^2}{2|\Delta\tau_{X,i}-\Delta\tau_{Y,j}| + s_{i}^2 + s_{j}^2}
    \\
    \beta_{XY,ij} & = & \frac{\sigma_u |\Delta\tau_{X,i}-\Delta\tau_{Y,j}|}{\sqrt{4\tau-2\Delta\tau_{X,i}-2\Delta\tau_{Y,j}+\sigma_{i}^2+\sigma_{j}^2}}
    \\
    \gamma_{XY,ij} & = & 
    \frac{\sigma_u |\Delta\tau_{X,i}-\Delta\tau_{Y,j}|}{\sqrt{2|\Delta\tau_{X,i}-\Delta\tau_{Y,j}| + s_{i}^2+s_{j}^2}}.
\end{eqnarray}
To compute the variances $\sigma_{S_X}$ and $\sigma_{S_Y}$, note that we must have $\Xi_{XY} = 1$ for $X=Y$, and therefore
\begin{equation}
    \sigma_{S_X}^2 = \frac{1+\sigma_w^2}{8\pi}
    \sum_{i=1}^{N_{\mathrm{ch}}}
    \sum_{j=1}^{N_{\mathrm{ch}}}
    f_{X,i} f_{X,j}
    \left(
    \ln \alpha_{XX,ij}
    +
    2 \sqrt{\frac{\beta_{XX,ij}^2}{1+\beta_{XX,ij}^2}}\sinh^{-1}\beta_{XX,ij}
    -
    2 \sqrt{\frac{\gamma_{XX,ij}^2}{1+\gamma_{XX,ij}^2}}\sinh^{-1}\gamma_{XX,ij}
    \right)
    \label{eq:variance}
\end{equation}
and similarly for $\sigma_{S_Y}^2$.

\section{\added{Sensitivity of the results to parameter choices}}
\label{app:parameter_sensitivity}

\begin{figure}
    \includegraphics[width=\columnwidth]{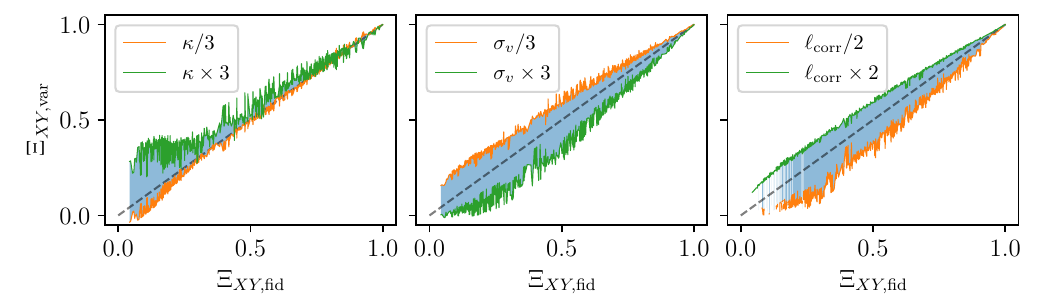}
    \caption{\added{Sensitivity of predicted elemental cross-correlations to variations in parameter choices. In each panel, the horizontal coordinate $\Xi_{XY,\mathrm{fid}}$ is the predicted cross-correlation for an element pair $XY$ using our fiducial parameter choices from \autoref{tab:parameters}, while the blue shaded vertical band shows the range of predictions resulting from varying the diffusion coefficient $\kappa$ by a factor of three (left panel), the velocity dispersion at birth $\sigma_v$ by a factor of three (middle panel), and the correlation length $\ell_\mathrm{corr}$ by a factor of two (right panel). The dashed black lines show the one-to-one relation, and the green and orange lines are colored differently to indicate which direction of variation corresponds to increasing versus decreasing the parameter being varied.\\
    \label{fig:Xi_var}
    }}
\end{figure}

\added{
Here we explore the sensitivity of our results to our choice of parameters. Our predicted cross-correlations depends primarily on four factors. One of these is the yields and delay times of the individual elements; these are significantly uncertain, but a review of the plausible range of variation in them is beyond the scope of this paper. The other three are the diffusion coefficient $\kappa$, the velocity dispersion for stars leaving their birth sites $\sigma_v$, and the correlation length $\ell_\mathrm{corr}$. The first of these is perhaps the most uncertain, since, as discussed in \autoref{ssec:model_basics}, even the description of elemental mixing as a diffusion process is necessarily approximate; we consider variations in this parameter by a factor of three entirely plausible. By contrast, the velocity dispersions of stellar drift cannot plausibly be more than a factor of $\approx 3$ larger than our fiducial choice of 2 km s$^{-1}$, since a factor of three larger than this (6 km s$^{-1}$) is comparable to the observed velocity dispersion of cold ISM \citep[e.g.,][]{Ianjamasimanana12a, Ianjamasimanana15a, Caldu-Primo13a, Caldu-Primo15a}. We therefore consider a factor of three variation in this parameter as well. Finally, as noted in \autoref{ssec:crosscorr_table}, $\ell_\mathrm{corr}$ is reasonably well-constrained by observations, and the scatter for Milky Way-like galaxies is observed to be roughly a factor of two; we therefore consider varying this parameter by a factor of two.
}

\added{
In \autoref{fig:Xi_var} we show the results of varying each of these parameters. To perform this experiment, we vary one of the the three parameters over the stated range while holding the other two constant, and recompute our predicted true cross-correlation for each element pair $XY$, not including any observational uncertainties. In each panel, the horizontal coordination shows our estimate of the cross-correlation for each element pair computed using our fiducial parameters, while the shaded vertical band shows the range of predictions we generate by varying one of these parameters within our stated uncertainty range.\footnote{\added{Note that for $\ell_\mathrm{core}$ some of the shading and results for $\ell_\mathrm{corr}/2$ are missing. This is because for certain elements with very long delay times, using $\ell_\mathrm{corr} = 500$ pc corresponds to a choice of $t$ that violates our assumption that $t$ is larger than the delay time for any element. Our model breaks down for these cases, and thus we omit them.}}
}

\added{
The figure makes clear that varying $\kappa$ by a factor of three has minimal effects on our predictions. The median variation in predicted $\Xi_{XY}$ over all element pairs is below 10\% (formally, the median value of $\exp[\ln(\Xi_{XY,\mathrm{max}} / \Xi_{XY,\mathrm{min}})/2]$ is  1.06, where $\Xi_{XY,\mathrm{min/max}}$ are the minimum and maximum values of $\Xi_{XY}$ that we predict as a result of varying $\kappa$), though it is larger for some elements with small correlation lengths, and the sensitivity to $\kappa$ is quite variable from one element pair to another depending on the particular nucleosynthetic pathways that contribute to the two elements in question. For most but not all elements an increase in $\kappa$ leads to an increase in the correlation. In general AGB-dominated elements are more sensitive to the choice of $\kappa$ than CCSN or TSN elements. The effects of varying $\sigma_v$ or $\ell_\mathrm{corr}$ are somewhat larger but still fairly modest: 20 - 30\%. Higher drift velocities, not surprisingly, tend to reduce elemental cross-correlations, while higher correlation lengths increase them. The overall conclusion to draw from our experiment, however, is that our qualitative conclusions are unchanged over plausible levels of variation in our parameters.
}

\bibliographystyle{aasjournal}
\bibliography{refs}

\end{document}